\documentclass[a4paper,11pt]{article}
\usepackage{amsmath,amssymb,amsfonts}
\usepackage[multiple,stable]{footmisc}
\usepackage[amsmath,amsthm,thmmarks]{ntheorem}
\allowdisplaybreaks[4]
\usepackage[noconfig]{refstyle}
\usepackage[dvipsnames]{xcolor}
\usepackage[hyperfootnotes=false, linktocpage=true, colorlinks, citecolor=blue, linkcolor=blue, urlcolor=Maroon]{hyperref}
\usepackage{array,tabularx,booktabs,geometry,subfigure,graphicx,longtable}
\usepackage[bf]{caption}
\usepackage{tikz}
\usepackage[all]{xy}
\usetikzlibrary{shapes,arrows}
\tikzstyle{block} = [rectangle, draw, text width=7em, text centered, rounded corners, minimum height=3em]
\usepackage{graphicx,color}
\usepackage{cite}
\usepackage{arydshln}
\usepackage{tikz-cd}

\let\eqref=\relax
\newref{eq}{name={eq.~},Name={Eq.~},names={eqs.~},Names={Eqs.~},rngtxt={-},refcmd=(\ref{#1})}
\newref{f}{name={footnote~},Name={Footnote~},names={footnotes~},Names={Footnotes~}}
\newref{app}{name={appendix~},Name={Appendix~},names={appendixes~},Names={Appendixes~}}
\newref{tab}{name={table~},Name={Table~},names={tables~},Names={Tables~}}
\newref{ch}{name={chapter~},Name={Chapter~},names={chapters~},Names={Chapters~}}
\newref{sec}{name={section~},Name={Section~},names={sections~},Names={Sections~}}
\newref{fig}{name={figure~},Name={Figure~},names={figures~},Names={Figures~}}
\numberwithin{equation}{section}
\newcommand{\eref}[1]{(\ref{#1})}

\newcommand{\eeq}{\end{equation}}
\newcommand{\beq}{\begin{equation}}
\newcommand{\ba}{\begin{array}}
\newcommand{\ea}{\end{array}}
\newcommand{\cB}{{\cal B}}

\newcommand{\cO}{{\cal O}}

\newcommand{\IP}{\mathbb P}

\newcommand{\be}{\begin{equation}}
\newcommand{\ee}{\end{equation}}
\newcommand{\bea}{\begin{equation}\begin{aligned}}	% note: abbreviations for \begin{align} and \end{align} don't work!
\newcommand{\eea}{\end{aligned}\end{equation}}		% note: \begin{equation}\begin{split}... produces pdf/hyperref warnings:
\newcommand{\iddots}{\mathinner{\mkern2mu\raise1pt\hbox{.}\mkern2mu \raise4pt\hbox{.}\mkern2mu\raise7pt\hbox{.}\mkern1mu}}

\providecommand{\id}{\leavevmode\hbox{\small$\mathrm{1}$\kern-3.8pt\normalsize$\mathrm{1}$}}
\def\fnote#1#2{\begingroup\def\thefootnote{#1}\footnote{#2}
     \addtocounter{footnote}{-1}\endgroup}
\begin{document}

\vspace{1cm}

\title{
       \vskip 40pt
       {\huge \bf Heterotic/Heterotic and Heterotic/F-theory Duality }}

\vspace{2cm}

\author{Lara B. Anderson${}^{1,2}$, He Feng${}^{1}$, Xin Gao${}^{3}$ and Mohsen Karkheiran${}^{1,2}$}
\date{}
\maketitle
\begin{center} {\small ${}^1${\it Physics Department, Robeson Hall, Virginia Tech, Blacksburg, VA 24061, USA}\\
${}^{2}$ {\it Simons Center for Geometry and Physics,~Stony Brook,~NY 11794,~USA} \\
${}^3${\it Dipartimento di Fisica, Universita di Roma \lq\lq Tor Vergata",  and \\
I.N.F.N. Sezione di Roma  \lq\lq Tor Vergata",  Rome, 00133, Italy}
}\\
\fnote{}{lara.anderson@vt.edu, fenghe@vt.edu, xingao@roma2.infn.it, mohsenka@vt.edu}
\end{center}

\begin{abstract}
\noindent
We consider heterotic target space dual (0,2) GLSMs on elliptically fibered Calabi-Yau manifolds. In this context, each half of the ``dual" heterotic theories must in turn have an F-theory dual. Moreover, the apparent relationship between two heterotic compactifications seen in (0,2) heterotic target space dual pairs should, in principle, induce some putative correspondence between the dual F-theory geometries. It has previously been conjectured in the literature that (0,2) target space duality might manifest in F-theory as multiple $K3$-fibrations of the same elliptically fibered Calabi-Yau manifold. In this work we investigate this conjecture in the context of both 6-dimensional and 4-dimensional effective theories and demonstrate that in general, (0,2) target space duality cannot be explained by such a simple phenomenon alone. In all cases, we provide evidence that non-geometric data in F-theory must play at least some role in the induced F-theory correspondence, while leaving the full determination of the putative new F-theory duality to future work.
\end{abstract}

\thispagestyle{empty}
\setcounter{page}{0}
\newpage

\tableofcontents

%\subfile{introduction}

%%%%%%%%%%%%%%%%%%%%%%%%%%%%%%%%%%%%%

\section{Introduction}\label{sec:intro}

%%%%%%%%%%%%%%%%%%%%%%%%%%%%%%%%%%%%%
Heterotic target space duality was first observed in \cite{Distler:1995bc} and further explored in \cite{Blumenhagen:1997vt,Blumenhagen:1997cn,Blumenhagen:2011sq,Rahn:2011jw,Anderson:2016byt}. The basic premise is simple to state: two distinct $(0,2)$ GLSMs sharing a non-geometric (i.e. Landau-Ginzburg or Hybrid) phase can be found to have apparently identical $4$-dimensional, $\mathcal{N}=1$ target space theories. In these cases, the GLSMs are distinct and the geometric phases of the two theories lead to manifestly different Calabi-Yau manifolds and vector bundles over them. However, the ensuing 4-dimensional theories, arising as large volume compactifications of the $E_8 \times E_8$ heterotic string, contain at least the same gauge symmetry and 4-dimensional massless particle spectrum. Although not yet understood as a true string duality, this phenomenon has been referred to as \emph{(0,2) Target Space Duality} (TSD) \cite{Distler:1995bc}. A more recent ``landscape" survey of such theories \cite{Blumenhagen:2011sq,Rahn:2011jw} showed that it is not just in special cases that such dualities can occur, but rather the vast majority of $(0,2)$ GLSMs contain non-geometric phases that can be linked to other $(0,2)$ GLSMs in this way. Moreover, recent work \cite{Anderson:2016byt} demonstrated that in some cases TSD also seems to preserve the form of non-trivial D- and F-term potentials of the 4-dimensional theory to a remarkable degree.

In the present work we aim to further explore the consequences of target space duality in the context of yet another duality -- that between heterotic string compactifications and F-theory. As has been observed since the first investigations into TSD \cite{Chiang:1997kt,Blumenhagen:1997vt}, this non-trivial duality of distinct heterotic backgrounds could potentially also lead to an entirely new duality structure within F-theory. Since heterotic and F-theory vacua consist of two of the most promising frameworks for string model building within $4$-dimensional string compactifications, it make sense to search for such novel and unexplored dualities to better understand redundancies within the space of such theories. In addition, if new dualities exist, they could also provide deep insights into the structure of the effective physics, or perhaps even new computational tools (as has manifestly proved to be the case with mirror symmetry in Type II compactifications of string theory, see e.g. \cite{Hori:2003ic}).

Compactifications of the heterotic string and F-theory can lead to identical effective theories in the situation that the background geometries of the two theories both exhibit fibration structures \cite{Morrison:1996na}. Namely, heterotic string theory compactified on a Calabi-Yau n-fold with an elliptic fibration
\beq\label{het_background}
\pi_h: X_n \stackrel{\mathbb{E}}{\longrightarrow} B_{n-1}
\eeq
over a base manifold $B_{n-1}$, leads to the same effective physics as F-theory compactified on a Calabi-Yau $(n+1)$-fold with a $K3$ fibration over the \emph{same} base manifold, $B_{n-1}$:
\beq
\pi_f: Y_{n+1} \stackrel{K3}{\longrightarrow} B_{n-1}
\eeq
In order to have a well-defined F-theory background, the $(n+1)$-fold $Y_{n+1}$ must also be elliptically fibered, with compatible elliptic/K3 fibrations \cite{Vafa:1996xn,Morrison:1996na}.

In the context of (potential) heterotic/heterotic dualities and heterotic/F-theory duality then, there are a number of natural questions that arise. Suppose that $(X_3, \pi: V\to X_3)$ and $(\tilde{X}_3, \pi: \tilde{V} \to \tilde{X}_3)$ are the requisite background geometries (i.e. (manifolds, vector bundles)) defining two TSD heterotic theories, then these questions include:
\begin{itemize}
\item Can Target Space Dual pairs be found in which both $X$ and $\tilde{X}$ are elliptically fibered as in \eref{het_background}? In principle, these two fibrations need not be related in any obvious way, for example: two topologically distinct CY 3-folds $\pi: X_3 \to B_{2}$ and $\tilde{\pi}: \tilde{X}_3 \to \tilde{B}_2$, with distinct 2 (complex) dimensional base manifolds $B_2$, $\tilde{B}_2$ to their fibrations.
\item If such elliptically fibered CY 3-fold geometries can be found within a TSD pair, this will in principle lead to two CY $4$-folds, $Y_4$ and $\tilde{Y}_4$, as dual backgrounds for F-theory. It should follow by construction that these two geometries lead to the same $4$-dimensional effective theory (or at least the same massless spectrum). How can this apparent duality be understood in the context of F-theory? How are $Y_4$ and $\tilde{Y}_4$ related?
\end{itemize}
For the first point, to our knowledge no explicit pairs of elliptically fibered TSD heterotic geometries have yet appeared in the literature. However, at least one proposal for the latter point has been posited. In \cite{Blumenhagen:2011sq} it was proposed that if fibered heterotic TSD pairs could be found, one possibility for the induced duality in F-theory would be the existence a CY $4$-fold with a single elliptic fibration, but \emph{more than one K3 fibration}:

\begin{displaymath}
 \xymatrix{
& Y_4 \ar[ld]_{K3}^{\pi_f} \ar[rd]^{\widetilde{K3}}_{\tilde \pi_f} &\\
{B}_{2}& & \widetilde{B_{2}}},
\end{displaymath}
where each fibration can be seen as the F-theory dual of one of the heterotic vacua (associated to $(X,V)$ or $(\tilde{X}, \tilde{V})$ respectively). Since by its very definition, F-theory requires that $Y_4$ is also elliptically fibered, this would imply that each $K3$-fiber appearing above is itself also elliptically fibered. Moreover, since the elliptic fibration of F-theory that determines the effective physics (i.e. gauge symmetry, matter spectrum, etc), in order for the two $K3$ fibrations to lead to identical effective theories, it would be expected that in fact in this scenario, $Y_4$ has only one elliptic fibration, but that it is compatible with two distinct $K3$ fibrations. If these compatible fibration structures were to exist it must be that the base of the elliptic fibration, $\rho: Y_{4} \to B_3$, must have two different $\mathbb{P}^1$-fibrations:
\begin{displaymath}
 \xymatrix{
& Y_4 \ar[d]^{\mathbb E}_{\rho_f} & \\
& B_3 \ar[ld]_{\mathbb P^1}^{\sigma_f} \ar[rd]^{\mathbb P^1}_{\tilde \sigma_f} &\\
B_2& & \widetilde B_2}
\end{displaymath}

The scenario above is one obvious way in which a ``duality" of sorts could arise in F-theory. Of course, in this case the essential F-theory geometry is not changing, only the $K3$ fibrations which determine the heterotic dual. This is clearly not the only possibility. As one alternative, it could prove that the F-theory duals of heterotic TSD pairs are in fact two distinct CY 4-folds, $Y_4$ and $\tilde{Y}_4$ whose gauge symmetries, massless spectra and effective ${\cal N}=1$ potentials are ultimately the same through non-trivial G-flux in the background geometry. We can summarize these two options for the induced duality in F-theory as follows
\begin{enumerate}
\item (Possibility 1) Heterotic TSD $\Leftrightarrow$ Multiple $K3$ fibrations in a single F-theory geometry (and hence manifestly leading to the same effective physics).
\item (Possibility 2) Heterotic TSD $\Leftrightarrow$ Two distinct pairs of manifolds and G-flux, $(Y_4, G_4)$ and $(\tilde{Y}_4, \tilde{G_4})$ which lead to the same effective physics in F-theory.
\end{enumerate}

In this work we investigate the two questions listed above and provide explicit examples of heterotic target space dual pairs with the requisite fibration structures to lead to F-theory dual theories. As we will outline in the following sections, substantial technical difficulties arise in explicitly computing the full F-theory duals of these heterotic theories. In the present work we do not attempt to fully determine these dual F-theories and instead provide evidence for our primary conclusion: \emph{multiple fibrations in F-theory cannot in general explain the dual physics of $(0,2)$ TSD}.

To overcome some of the technical obstacles of heterotic/F-theory duality, we begin our analysis by actually considering heterotic/F-theory dual pairs in $6$-dimensional effective theories rather than in $4$-dimensions. In this context, the heterotic duality is a trivial one -- TSD pairs simply generate two bundles over K3 with the same second Chern class, and are thus trivially guaranteed to give rise to the same massless spectrum (see e.g. \cite{Bershadsky:1996nh}). However, this very simple framework for heterotic TSD pairs allows us to explicitly perform Fourier-Mukai transforms to render the data of a holomorphic, stable vector bundle over $K3$ into its spectral cover \cite{FMW}. With this data, we are able to explicitly construct examples of F-theory duals and verify that in fact they cannot arise as multiple $K3$ fibrations of a CY 3-fold, $Y_3$, determining an F-theory background. The results of this study are presented in Section \ref{6D_Section}.

Turning once more to our primary area of interest in ${\cal N}=1$ and heterotic compactifications on CY 3-folds, we outline the essential ingredients determining the dual F-theory geometry. We find that in general a number of technical tools are missing for fully determining the F-theory physics. Some of these we have developed and will appear separately \cite{us_mohsen,us_conicbundles} while others we leave to future work. However, we are able to indicate that in general the intermediate Jacobians of the dual F-theory geometries must play some role in the new ``F-theory duality" whatever it may prove to be. This leads to the presence of essential data not associated to the complex structure of the CY 4-fold alone, but G-flux as well. In the singular limit such fluxes are well known to have the potential to dramatically change the effective physics through so-called T-brane solutions \cite{Cecotti:2010bp,Anderson:2013rka,Anderson:2017rpr} and other possibilities.

 In the following sections we will explore these ideas in detail. In particular, the paper is organized in the following way. In Section \ref{Background} we review briefly the essential aspects of $(0,2)$ Target Space Duality. In Section \ref{sec:CICY}, we provide the first non-trivial examples to appear in the literature of heterotic TSD pairs in which both CY 3-folds, $X$ and $\tilde{X}$ are elliptically fibered. In these cases, the heterotic geometries are smooth (consisting of smooth so-called CICY threefolds \cite{Candelas:1987kf} and stable, holomorphic vector bundles defined via the monad construction \cite{horrocks} over them) and lead to well controlled, perturbative heterotic theories. However we will demonstrate in this and subsequent sections that existing techniques in the literature to determine dual F-theory geometries, as outlined in Section \ref{4D_thoughts} are insufficient to determine the geometry of $Y_4$ and $\tilde{Y}_4$ in these cases. However, we none-the-less still find some evidence indicating that multiple fibrations of $Y_4$ cannot be the F-theory manifestation of $(0,2)$ TSD. 

To make concrete the dual F-theory geometry we move to $6$-dimensional examples in Section \ref{6D_Section}. More precisely, we consider heterotic TSD theories consisting of pairs of bundles over $K3$ in which the second Chern class of both $V$ and $\tilde{V}$ is taken to be 12. In this case, it is possible that the F-theory geometry, $Y_3$ is multiply fibered as described above. However after finding the spectral data (i.e. Fourier Mukai transform) of these bundles, we can explicitly construct the dual F-theory geometry and find that it does not in general agree with what can be obtained by multiple fibrations. We will argue further that the F-theory ``image" of target space duality under the Heterotic/F-theory map shouldn't be purely geometric even in $6$-dimensions, but rather it can be related to the intermediate Jacobian of the CY 3-fold. With these tools and observations in hand we return to the F-theory duals of four dimensional, ${\cal N}=1$ heterotic theories in Section \ref{4D_thoughts}.

Finally, in the Appendices we consider a handful of examples illustrating both the range of possibilities arising in heterotic TSD dual geometries, as well as potential pitfalls that can arise in constructing dual pairs.

%\subfile{background}

%%%%%%%%%%%%%%%%%%%%%%%%%%%%%%%%%%%%%

\section{A brief review of $(0,2)$ Target Space Duality}\label{Background}

%%%%%%%%%%%%%%%%%%%%%%%%%%%%%%%%%%%%%

Heterotic target space duality is best understood in the context of heterotic string compactifications associated to $(0, 2)$ gauged linear sigma models (GLSMs). It was first observed by Distler and Kachru in 1995\cite{Distler:1995bc}, and further studied by Blumenhagen \cite{Blumenhagen:1997vt,Blumenhagen:1997cn} with later a landscape study \cite{Blumenhagen:2011sq}. 
%In 2011 a landscape study \cite{Blumenhagen:2011sq} reveals some mystery of  target space duality.  
The GLSM provides a description of the complexified compact stringy K\"ahler moduli space which is divided into various phases \cite{Witten:1993yc}. The freedom to vary a Fayet-Illiopolos parameter links variety of distinct phases including the geometric phases (associated to target space geometries like Calabi-Yau threefolds $X$ and holomorphic vector bundle $V$), non-geometric phase (commonly a Landau-Ginzburg phase), and a rich variety of hybrid phase.
Described in the (0,2) GLSM language, target space duality is realized by exchanging two certain types of charges in theory, which is defined by $(X, V)$ in the geometric phase,  to give a different configuration $(\tilde X, \tilde V)$ from the original one  while leaving the superpotential invariant and sharing a common Landau-Ginzburg phase. Meanwhile,  in the geometric phases, this pair of theories $(X,  V)$ and $(\tilde X, \tilde V)$ %However, the full understanding of this duality is still missing,
%until now we still don't have a full understanding of it, 
  preserve net number of moduli, the complete charged and singlet particle spectra.  
 
In an Abelian GLSM,  there exists multiple $U(1)$ gauge fields $A^{(\alpha)}$ with $\alpha = 1,...,r$, two sets of chiral superfields as $\{X_i|i=1,...,d\}$ with $U(1)$ charges $Q_i^{(\alpha)}$, and $\{P_l|l=1,...,\gamma\}$ with $U(1)$ charges $-M_l^{(\alpha)}$. Furthermore, there are two sets of Fermi superfields: $\{\Lambda^a|a=1,...,\delta\}$ with charges $N_a^{(\alpha)}$, and $\{\Gamma^j|j=1,...,c\}$ with charges $-S_j^{(\alpha)}$.  
These charges are given in order to realize the Calabi-Yau manifolds as complete intersection hypersurfaces in ambient space (Complete Intersection Calabi-Yau (CICY)) and stable, holomorphic vector bundles over them in some geometric phase. 
As a result, we will require the charges $Q_i^{(\alpha)} \geq 0$ and  for each $i$, there exists at least one $r$ such that $Q_i^{(\alpha)}>0$. Similar assumption of (semi-)positivity will also hold for the charges $S_j^{(\alpha)}$ and $M_l^{(\alpha)}$. However, in some cases we will consider solutions in which charges $N_a^{(\alpha)}$ may be negative.
Then the  field content and charges of GLSM can be summarized in the following "charge matrix", 
\begin{equation}
\begin{aligned}
&\begin{array}{|c||c|}
\hline
x_i & \Gamma^j \\
\noalign{\hrule height 1pt}
\begin{array}{cccc}
 Q^{(1)}_1 & Q^{(1)}_2 & \ldots & Q^{(1)}_d\\
 Q^{(2)}_1 & Q^{(2)}_2 & \ldots & Q^{(2)}_d\\
 \vdots  &    \vdots  &  \ddots  &  \vdots  \\  
 Q^{(r)}_1 & Q^{(r)}_2 & \ldots & Q^{(r)}_d
\end{array}
&
\begin{array}{cccc}
-S^{(1)}_1 & -S^{(1)}_2 & \ldots & S^{(1)}_c\\
-S^{(2)}_1 & -S^{(2)}_2 & \ldots & S^{(2)}_c\\
 \vdots  &    \vdots  &  \ddots  &  \vdots  \\  
-S^{(r)}_1 & -S^{(r)}_2 & \ldots & S^{(r)}_c\\
\end{array}\\
\hline
\end{array}
\\[0.1cm]
&\begin{array}{|c||c|}
\hline
\Lambda^a & p_l \\
\noalign{\hrule height 1pt}
\begin{array}{cccc}
 N^{(1)}_1 & N^{(1)}_2 & \ldots & N^{(1)}_\delta\\
 N^{(2)}_1 & N^{(2)}_2 & \ldots & N^{(2)}_\delta\\
 \vdots  &    \vdots  &  \ddots  &  \vdots  \\  
N^{(r)}_1 & N^{(r)}_2 & \ldots & N^{(r)}_\delta
\end{array}
&
\begin{array}{cccc}
-M^{(1)}_1 & -M^{(1)}_2 & \ldots & -M^{(1)}_\gamma\\
-M^{(2)}_1 & -M^{(2)}_2 & \ldots & -M^{(2)}_\gamma\\
 \vdots  &    \vdots  &  \ddots  &  \vdots  \\ 
-M^{(r)}_1 & -M^{(r)}_2 & \ldots & -M^{(r)}_\gamma\\
\end{array}\\
\hline
\end{array}
\end{aligned}\label{charge_matrix}
\end{equation}
We can denote such starting point in the geometric phase as:
\beq\label{starting_point}
V_{N_1, \ldots N_{\delta}}[M_1, \ldots, M_{\gamma}] \longrightarrow \mathbb{P}_{Q_1, \ldots Q_d}[S_1, \ldots, S_c]
\eeq
Here anomaly cancellation condition requires the following linear and quadratic constraints for all $\alpha, \beta = 1, ..., r$:
\begin{eqnarray}\nonumber \label{anomalies}
\sum_{a=1}^\delta N_a^{(\alpha)} = \sum_{l=1}^\gamma M_l^{(\alpha)} \qquad&& \sum_{i=1}^d Q_i^{(\alpha)} = \sum_{j=1}^c S_j^{(\alpha)}\\
\sum_{l=1}^\gamma M_l^{(\alpha)} M_l^{(\beta)} - \sum_{a=1}^\delta N_a^{(\alpha)} N_a^{(\beta)} &=& \sum_{j=1}^c S_j^{(\alpha)} S_j^{(\beta)} - \sum_{i=1}^d Q_i^{(\alpha)} Q_i^{(\beta)}
\end{eqnarray}
GLSM is further described by a superpotential and a scalar potential, while the scalar potential has contributions from F-term and D-term:
\begin{eqnarray} \label{potential} \nonumber
S &=& \int d^2z d\theta \bigg[\sum_j \Gamma^j G_j (x_i) + \sum_{l,a} P_l \Lambda^a F_a^l (x_i) \bigg], \\
V_F &=& \sum_j \big |G_j(x_i) \big|^2 + \sum_a\big |\sum_l p_lF_a^l(x_i) \big|^2,\\ \nonumber
V_D &=& \sum_{\alpha=1}^r \bigg( \sum_{i=1}^d Q_i^{(\alpha)} |x_i|^2 - \sum_{l=1}^\gamma M_l^{(\alpha)}|p_l|^2 - \xi^{(\alpha)} \bigg)^2,
\end{eqnarray}
where the functions $G_j$ and $F_a^l$ are quasi-homogeneous polynomials with degrees shown in the following matrix:
\begin{equation}
%\begin{aligned}
\begin{array}{|c|}
\hline
 G^j \\
\noalign{\hrule height 1pt}
\begin{array}{cccc}
 S_1 & S_2 & \ldots & S_c
\end{array}\\
\hline
\end{array}
\quad
\begin{array}{|c|}
\hline
F_a{}^l  \\
\noalign{\hrule height 1pt}
\begin{array}{cccc}
 M_1-N_1 & M_1-N_2 & \ldots & M_1-N_\delta\\
 M_2-N_1 & M_2-N_2 & \ldots & M_2-N_\delta\\
 \vdots  &    \vdots  &  \ddots  &  \vdots  \\  
M_\gamma-N_1 & M_\gamma-N_2 & \ldots & M_\gamma-N_\delta
\end{array}\\
\hline
\end{array}
%\end{aligned}
\end{equation}
Furthermore, the function $F_a^l$ will be chosen to satisfy a {transversality condition} such that $F_a^l(x)=0$ only when $x_i=0$ for $i = 1, ..., d$.

The $\xi^{(\alpha)} \in \mathbb R$ in the D-term potential is the Fayet-Iliopoulos (FI) parameter which determines the structure of the vacuum. From the original Witten's paper\cite{Witten:1993yc}, consider the simple case with only one $U(1)$ so there is only one $\xi$: If $\xi>0$ then it's the geometric phase, described by a vector bundle over a Calabi-Yau manifold: $V_{N_1,...,N_\delta}[M_1,...,M_\gamma]\longrightarrow X \equiv \IP_{Q_1,...,Q_d}[S_1,...,S_c]$, 
where the Calabi-Yau manifold is defined by complete intersection hypersurfaces  in weighted projective space, i.e, CICY $X=\cap_{j=1}^{c} G_j$ with $G_j(x_i) = 0$, and the vector bundle is defined by 
\bea
V = \frac{ker(F_a^l)}{im(E_i^a)}
\eea
 with rank $rk(V) = (\delta - \gamma - r_{\cal V})$  through the monad on $X$:
\begin{equation}\label{full_monad}
0 \to \mathcal O_{X}^{\oplus r_{\mathcal V}} \xrightarrow{E_i^a} \bigoplus_{a=1}^\delta \mathcal O_{X}(N_a) \xrightarrow{F_a^l} \bigoplus_{l=1}^\gamma \mathcal O_{X}(M_l) \to 0.
\end{equation}

%For $\xi > 0$ the D-term implies that not all $X_i$ are zero simultaneously, thus not all $F_a$ are zero, and the F-term implies $G_j(x_i) = 0$ and $<p> = 0$. This ``geometric" phase contains the data of a $(0,2)$ non-linear sigma-model on a generally singular complete intersection, $X$, in a weighted projective space $\mathbb P_{Q_1,...,Q_d}[S_1,...,S_c]$. Moreover, the superpotential in \eref{superpotential} leads to a mass term of the form $\sum_a \pi \lambda^a F_a$, which makes massive one linear combination of the $\lambda^a$. 

 If $\xi<0$ then it's the Landau-Ginzburg phase described by a superpotential:
 \begin{equation}\label{LG_superpotential}
\mathcal W(x_i, \Lambda^a, \Gamma^i) = \sum_j\Gamma^jG_j(x_i) + \sum_a \Lambda^aF_a(x_i), 
\end{equation}
where it would be a  hybrid-type non-geometric phases if there  are multiple $U(1)'s$.

Now let us move to the target space duality.
The first observation is an exchange/relabeling of $G_j$'s and  $F_a$'s  will leave the superpotential  (\ref{LG_superpotential}) invariant. This observation indicates that two distinct GLSMs could ``share" a non-geometric phase in which the original  role of  $G_j$ and $F_a$ is obscured. 
So the full procedure of target space dual would be starting from a geometric phase, go to a Landau-Ginzburg phase, do a rescaling/relabeling of the fields, and go back to the geometric phase to get a new Calabi-Yau/vector bundle configuration.

If the Landau-Ginzburg phase exists, then the rescaling procedure is as follows, for a non-vanishing $p_l$ and all $i=1, \ldots k$:
\beq
\label{field_redefine}
\tilde{\Lambda}^{a_i}:=\frac{\Gamma^{j_i}}{\langle p_l \rangle} ~,~~~~\tilde{\Gamma}^{j_i}=\langle p_l \rangle \Lambda^{a_i}
\eeq
\beq
|| \tilde{\Lambda}^{a_i} ||= || \Gamma^{j_i} || - ||P_l ||~~~~,~~~|| \tilde{\Gamma}^{j_i} ||= || \Lambda^{a_i} || + || P_l ||,
\eeq
with  $\sum_i ||G_{j_i} || = \sum_i || F_{a_i}{}^{l} ||$ for anomaly cancellation. One thing to notice is that exchanging only one $F$ with one $G$ does nothing.  So in all examples two or more $F_a$'s are exchanged with two or more $G_j$'s.   It is clear that the ``relabeling" of fields at the shared Landau-Ginzburg point can mix the degrees of freedom in $h^{2,1}(X)$ and $h^1(X, End_0(V))$ in the target space dual. 
In the landscape \cite{Blumenhagen:2011sq}, the dual sides match in the number of charged matter, and the total number of massless gauge singlets, where the individual number of complex, K\"ahler and bundle moduli are interchanged as:
\begin{eqnarray}\nonumber
&&h^*(X,\wedge^{k}V) = h^*(\widetilde X, \wedge^{k} \widetilde V),  \quad\quad k = 1,2,\cdots , rk(V)\\
&&h^{2,1}(X) + h^{1,1}(X) + h^1_X(End_0(V)) = h^{2,1}(\widetilde X) + h^{1,1}(\widetilde X) +  h^1_X(End_0(\widetilde V))
\end{eqnarray}

Furthermore, 
more general target space duality are possible such that it can \emph{also} change the dimension of $h^{1,1}(\tilde X)$.  For example, if there is only one column in $G$, which is not enough to make the exchange, then a blow up of $\mathbb P^1$ on the manifold will help. This procedure leads to a dual models with an additional U(1) action. In this case,  it is necessary to re-write the initial GLSM in an equivalent/redundant way.  It is always possible to introduce into the GLSM a new coordinate (i.e. a new Fermi superfield) $y_1$ with multi-degree ${\mathcal B}$ and a new hypersurface (i.e. a Chiral superfield with opposite charge to the new Fermi superfield) $G^{\cB}$ corresponding to a homogeneous polynomial of multi-degree ${ \cB}$. Similar to (\ref {starting_point}), the above addition can be written
\beq\label{same_base}
V_{N_1, \ldots N_{\delta}}[M_1, \ldots, M_{\gamma}] \longrightarrow \mathbb{P}_{Q_1, \ldots Q_d,{\cal B}}[S_1, \ldots, S_c,{\cal B}].
\eeq
and the matrix form of such intermediate step should be:
\bea
\label{intermediate}
%\begin{center}
&\begin{tabular}{|ccccc||cccc| }
  \hline                       
   $x_1$ & \ldots & $x_d$ & $y_1$ & $y_2$  &$\Gamma^1$ & \dots & $\Gamma^c$ & $\Gamma^B$ \\
 \noalign{\hrule height 1pt}
     0   & \ldots &   0   &   1   &    1       &    0     & \ldots &     0      &    $-1$     \\
   $Q_1$ & \ldots & $Q_d$ &   ${\cal B}$ &    0       & $-S_1$   & \ldots & $-S_c$     &     $-{\cal B}$      \\
  \hline  
\end{tabular}\\
&\begin{tabular}{|cccc||cccc| }
  \hline                       
   $\Lambda^1$ & $\Lambda^1$ & \ldots & $\Lambda^\delta$ & $p_1$ & $p_2$ & \ldots & $p_\gamma$\\
 \noalign{\hrule height 1pt}
      0   & 0   & \ldots & 0 & $-1$ & 0 & \ldots & 0  \\
   $N_1$ & $N_2$ & \ldots & $N_\delta $ & $-M_1$ & $-M_2$ &  \ldots & $-M_\gamma$  \\
  \hline  
\end{tabular}
%\hfill
%\end{center}
\eea
Suppose that in an example there are two chosen map elements $F_1^1$ and $F_2^1$ that have been chosen to be interchanged with a defining relation $S_1$. In this case we can choose the redundant new coordinate, $y_1$, to have charge
\begin{equation}
\label{addu1}
||{\cal B}|| = ||F_1^1|| + ||F_2^1|| - S_1
\end{equation}
For the initial configuration,  $||G_1|| + ||G_2|| = ||F_1^1|| + ||F_2^1||$ where $G_1,  G_2$ are $S_1, \cB$.
Under the re-labelings required in  (\ref{field_redefine}) one can choose,
\beq\label{u1_field_defs}
\tilde{N}_1=M_1-S_1 ~,~\tilde{N}_2=M_1 - {\cal B}~,~\tilde{S}_1 = ||F_1^1||~,~\tilde{{\cal B}}=||F_2^1 ||
\eeq
Applying the field redefinitions in \eref{u1_field_defs} we arrive at last to the new configuration
\bea
&\begin{tabular}{|ccccc||cccc| }
  \hline                       
   $x_1$ & \ldots & $x_d$ & $y_1$ & $y_2$  &$\tilde{\Gamma}^1$ & \dots & $\Gamma^c$ & $\tilde{\Gamma}^B$ \\
 \noalign{\hrule height 1pt}
     0   & \ldots &   0   &   1   &    1       &    $-1$     & \ldots &     0      &    $-1$     \\
   $Q_1$ & \ldots & $Q_d$ &   ${\cal B}$ &    0       & $-(M_1-N_1)$   & \ldots & $-S_c$     &     $-(M_1-N_2)$       \\
  \hline  
\end{tabular}\\
&\begin{tabular}{|cccc||cccc| }
  \hline                       
 $\tilde{\Lambda}^1$ & $\tilde{\Lambda}^1$ & \ldots & $\Lambda^\delta$ & $p_1$ & $p_2$ & \ldots & $p_\gamma$\\
 \noalign{\hrule height 1pt}
 1   & 0   & \ldots & 0 & $-1$ & 0 & \ldots & 0  \\
 $M_1-S_1$ & $M_1-{\cal B}$ & \ldots & $N_\delta $ & $-M_1$ & $-M_2$ &  \ldots & $-M_\gamma$  \\
  \hline  
\end{tabular}
\eea
%For simplicity, in this work all examples will be chosen so that the target space duals satisfy ${\cal %B}=0$ which will guarantee that the theories stay within the class of simple complete intersection %Calabi-Yau manifolds in products of projective spaces \cite{Candelas:1987kf,Hubsch:1986ny}.

In subsequent sections we will consider primarily examples of this latter kind in which all three types of singlet moduli - K\"ahler, complex structure, and bundle moduli -- are interchanged in the target space duality procedure. We turn to such an example next in which both $X$ and $\tilde{X}$ are elliptically fibered.

%\subfile{eg_CICY}

%%%%%%%%%%%%%%%%%%%%%%%%%%%%%%%%%%%%%

\section{A target space dual pair with elliptically fibered Calabi-Yau threefolds} 
\label{sec:CICY}

%%%%%%%%%%%%%%%%%%%%%%%%%%%%%%%%%%%%%

Before we can begin to investigate the consequences of $(0,2)$ target space duality for F-theory, a non-trivial first step is to establish if examples exist in which both halves of a TSD pair in turn lead to F-theory dual geometries. In this section we explicitly provide a first example of such a pair.

In the following example, we will find that the CY manifolds, $X$ and $\tilde{X}$, consist of two Complete Intersection Calabi-Yau 3-folds (so-called ``CICYs" \cite{Hubsch:1992nu, Candelas:1987kf}), each of which is fibered over (a different) complex surface $B_2$. These two CICY 3-folds are related by a conifold transition \cite{Hubsch:1992nu} and can be constructed via the target space duality algorithm in which an additional $U(1)$ symmetry is added to the dual GLSM as in Section \ref{Background}.

\subsection{A tangent bundle deformation}

To investigate these results, a simple starting point is given below -- a dual pair for which the Calabi-Yau manifolds, $X$ and $\tilde{X}$, are related by a conifold transition.
Consider the following CICY 3-fold, described by a so-called ``configuration matrix" \cite{Hubsch:1992nu}
\begin{equation}
X = \left[ \begin{array}{c|cc} \IP^1 & 1&1\\  \IP^2 & 1&2\\ \IP^2 &1&2\end{array} \right]~.
\end{equation}
Here the columns indicate the ambient space (a product of complex projective spaces) and the degrees of the defining equations in that space. The Hodge numbers are $h^{1,1}(X)=3$ and $h^{2,1}(X)=60$. Over this manifold, we choose a simple vector bundle built as a deformation of the holomorphic tangent bundle to $X$. In the present case we will choose this bundle to be a rank $6$ smoothing deformation of the reducible bundle
\beq
V_{red}={\cal O}^{\oplus 3} \oplus TX~.
\eeq
The smooth, indecomposable bundle will be defined\footnote{See \cite{Anderson:2008ex} for discussions of this deformation problem and local moduli space.} as a kernel $V \equiv ker(F_a^l)$ via the short exact sequence \begin{equation}
\label{short_monad}
0 \to V \to \bigoplus_{a=1}^\delta \mathcal O_{\mathcal M}(N_a) \xrightarrow{F_a^l} \bigoplus_{l=1}^\gamma \mathcal O_{\mathcal M}(M_l) \to 0.
\end{equation}
which is the simple case of (\ref{full_monad}) when  $E_i^{a}=0$. 

In the language of GLSM charge matrices, the manifold and rank 6  monad bundle $(X, V)$ are given by the following  charge matrix:
\begin{equation}\begin{aligned}&\begin{array}{|c|c| |c|c|}
\hline
x_i & \Gamma^j & \Lambda^a & p_l \\ \noalign{\hrule height 1pt}\begin{array}{cccccccc}
1& 1& 0& 0& 0& 0& 0& 0 \\
0& 0& 1& 1& 1& 0& 0& 0 \\
0& 0& 0& 0& 0& 1& 1& 1
\end{array}&\begin{array}{cc}
 -1& -1   \\
 -1& -2  \\
 -1& -2
\end{array}&\begin{array}{cccccccc}
1& 1& 0& 0& 0& 0& 0& 0 \\
0& 0& 1& 1& 1& 0& 0& 0 \\
0& 0& 0& 0& 0& 1& 1& 1
\end{array}&\begin{array}{cc}
 -1& -1   \\
 -1& -2  \\
 -1& -2
\end{array}\\
\hline
\end{array}\end{aligned}\label{eq:eg31}\end{equation}
%This is a CY 3-fold with rank 6 bundle.
The reason that this rank $6$ bundle makes for a particularly simple choice of gauge bundle is that in this case the GLSM charges associated to the manifold and the bundle are identical (as can be seen above).  As a result, anomaly cancellation conditions such as the requirement that
\beq
c_2(TX)=c_2(V)
\eeq
(realized as \eref{anomalies} in the GLSM) are automatically satisfied. 

Expanding the second Chern class of the manifold in a basis of $\{1,1\}$ forms $J_r$, $r=1, \ldots h^{1,1}$ we have
\bea
\label{c2tangent}
c_2(TX) = 3J_1J_2  +J_2^2 +3J_1J_3 +5J_2J_3 +J_3^2
\eea

Following the standard $(0,2)$ target space duality procedure, it is easy to produce the TSD geometry ($\widetilde X, \widetilde V$).  In this case, the duals we consider mix all three types of heterotic geometry moduli and induce an additional $U(1)$ gauge symmetry in the GLSM. 
As an intermediate step we form the equivalent GLSM charge matrix with an additional $U(1)$ outlined in Section \ref{Background}  (choosing $\cB=0$) and introduce a repeated entry in the monad bundle charges which does not change either the geometry of GLSM field theory\footnote{See \cite{Blumenhagen:2011sq} for details of this argument.} This leads us to the following charge matrix with a new $\IP^1$ row and a new column  $\Gamma^{\cal B}$ as in (\ref{intermediate}):
\begin{equation}\begin{aligned}&\begin{array}{|c|c| |c|c|}
\hline
x_i & \Gamma^j & \Lambda^a & p_l \\ \noalign{\hrule height 1pt}\begin{array}{cccccccccc}
1& 1& 0& 0& 0& 0& 0& 0& 0& 0 \\
0& 0& 1& 1& 1& 0& 0& 0& 0& 0 \\
0& 0& 0& 0& 0& 1& 1& 1& 0& 0 \\
0& 0& 0& 0& 0& 0& 0& 0& 1& 1 
\end{array}&\begin{array}{ccc}
 0& -1& -1   \\
 0& -1& -2  \\
 0& -1& -2 \\
 -1&  0&  0
\end{array}&\begin{array}{ccccccccc}
1& 1& 0& 0& 0& 0& 0& 0& 1 \\
0& 0& 1& 1& 1& 0& 0& 0& 2 \\
0& 0& 0& 0& 0& 1& 1& 1& 1 \\
0& 0& 0& 0& 0& 0& 0& 0& 0
\end{array}&\begin{array}{ccc}
 -1& -1& -1   \\
 -1& -2& -2  \\
 -1& -2& -1  \\
  0& -1&  0
\end{array}\\
\hline
\end{array}\end{aligned}\end{equation}

Finally, we can perform the field redefinitions in this intermediate geometry to obtain the final TSD. Here we choose two map elements -- in this case, $F_8^2$ and $F_9^2$ -- to be interchanged with a defining relation $G_2$ with degree $||S_2|| = \{1,2,2\}$.
Such a choice satisfies the linear anomaly cancelation (\ref{addu1})  since $||S_2||+ {\bf 0}  = ||F_8^2|| + ||F_9^2||$.  In the intermediate configuration, applying the field redefinitions (\ref{u1_field_defs}) gives:
\bea
 \tilde N_8 = M_2- S_2 = 0, \quad \tilde{N}_9=M_2, \quad   \tilde S_2 := F_9^2, \quad \tilde \cB := F_8^2.
\eea
This leads us at last to the dual charge matrix associated to ($\widetilde X, \widetilde V$) with $h^{1,1}(\tilde X)=4$ and $h^{2,1}(\tilde X)=60$:
\begin{equation}\begin{aligned}&\begin{array}{|c|c| |c|c|}
\hline
x_i & \Gamma^j & \Lambda^a & p_l \\ \noalign{\hrule height 1pt}\begin{array}{cccccccccc}
1& 1& 0& 0& 0& 0& 0& 0& 0& 0 \\
0& 0& 1& 1& 1& 0& 0& 0& 0& 0 \\
0& 0& 0& 0& 0& 1& 1& 1& 0& 0 \\
0& 0& 0& 0& 0& 0& 0& 0& 1& 1  
\end{array}&\begin{array}{ccc}
 0& -1& -1   \\
 0& -1& -2  \\
-1& -1& -1 \\
-1&  0& -1
\end{array}&\begin{array}{ccccccccc}
1& 1& 0& 0& 0& 0& 0& 0& 1 \\
0& 0& 1& 1& 1& 0& 0& 0& 2 \\
0& 0& 0& 0& 0& 1& 1& 0& 2 \\
0& 0& 0& 0& 0& 0& 0& 1& 0
\end{array}&\begin{array}{ccc}
 -1& -1& -1   \\
 -1& -2& -2  \\
 -1& -2& -1  \\
  0& -1&  0
\end{array}\\
\hline
\end{array}\end{aligned}\label{eq:eg31dual}\end{equation}
Again, in the new configurations the anomaly cancellation condition are satisfied (as was proved in general to happen in \cite{Blumenhagen:2011sq}). To make sure they are true target space duals, we will show that these two different geometric phases preserve the net multiplicities of charged matter, and the total number of massless gauge singlets, while the individual number of complex, K\"ahler and bundle moduli are changed.
First, it is clear that the low-energy gauge group $G$ in the 4-dimensional gauge theory is given by the commutant of the structure group, $H$ in $E_8 \times E_8$ of the bundles defined over the CY manifold. Here there is only one bundle (saturating the anomaly cancellation condition on $c_2(V)$). We choose to embed this structure group into one of the two $E_8$ factors and considering the other $E_8$ factor as an unbroken, hidden sector gauge symmetry. 

In order to find the matter field representations, the adjoint $\bf 248$ of $E_8$ must be decomposed under the subgroup $G \times H$. In the present case, the rank $6$ bundles with $c_1=0$ indicates the structure group $H= SU(6)$, which leads the charged matter spectrum can be determined by   the decomposition of $E_8$ into representations of the maximal subgroup $SU(2) \times SU(3) \times SU(6)$:
\begin{equation}
 {\bf 248}_{E_8}\rightarrow \left[({\bf 3},{\bf 1},{\bf 1})\oplus ({\bf 1},{\bf 8},{\bf 1})\oplus ({\bf 1},{\bf 1},{\bf 35})\oplus ({\bf 1},{\bf 3},\overline{\bf 15})
 \oplus ({\bf 1},\overline{\bf 3},{\bf 15})
 \oplus ({\bf 2},{\bf 3},{\bf 6})\oplus (\overline{\bf 2},\overline{\bf 3},\overline{\bf 6})\oplus ({\bf 2},{\bf 1},{\bf 20})\right]\label{248dec}
\end{equation} 
As a result, the multiplicity of fields in the 4-dimensional theory transforming in representations of  $SU(2)\times SU(3)$ is counted by those transforming in an $SU(6)$ representation over the CY. The latter are counted by the dimension of bundle valued cohomology groups, $H^*(X, \wedge^k V)$, for assorted values of $k$ (see \cite{Green:2012pqa,Anderson:2018pui} for details).

It is helpful to note that for a vector bundle $V$, on a Calabi-Yau 3-fold, $X$, the cohomology groups of the bundle and its dual are related by Serre duality as $H^{i}(X, V) = H^{3-i} (X, V^*)^*$ and when $H=SU(n)$, $H^*(X, \wedge^k V) \simeq H^*(X, \wedge^{n-k} V^*)$. Finally, a necessary condition for $\mu$-stability of the vector bundle $V$ is $h^0 (X, V) = 0$ which is satisfied for tangent bundle deformations considered here (by direct computation).

With these observations in hand, the multiplicity of the charged chiral matter spectrum of these dual pair theories can be determined by computing corresponding vector bundle valued cohomology classes on the Calabi-Yau 3-fold:
\begin{eqnarray} \nonumber
({\bf 2}, {\bf  3})'s: \qquad & h^1( V) = 57 & \qquad h^1( \widetilde V) = 57 \\ \nonumber
(\overline{\bf 2}, \overline{\bf  3})'s: \qquad & h^1( V^*) = 0 & \qquad h^1( \widetilde V^*) = 0 \\ 
({\bf 1}, {\bf  3})'s: \qquad & h^1(\wedge^2 V) = 115 & \qquad h^1(\wedge^2 \widetilde V) = 115 \\
({\bf 1}, \overline{\bf  3})'s: \qquad & h^1(\wedge^2 V^*) = 1 & \qquad h^1(\wedge^2 \widetilde V^*) = 1 \nonumber\\
({\bf 2},{\bf 1})'s: \qquad & h^1(\wedge^3 V) = 2& \qquad h^1(\wedge^3 \widetilde V) = 2  \nonumber
%({\bf 1}, {\bf 3})'s: \qquad & h^1(\wedge^4 V) = 1& \qquad h^1(\wedge^4 \widetilde V) = 1
\end{eqnarray}

Furthermore,  the low-energy theory has massless gauge singlets, $(\bf 1, 1)$, which are counted by $h^1(V \otimes V^*) = h^1({End_0(V)})$. 
There are additional singlets, beyond those related to the complex structure and K\"ahler deformations of the Calabi-Yau 3-fold, which are counted by $h^{2,1}(X)$ and $h^{1,1}(X)$.
The total number of singlet moduli are counted by:
\begin{eqnarray} \nonumber
 h^{1,1}(X)+h^{2,1}(X)+h^1(End_0(V))=3+60+292=355, \\
h^{1,1}(\widetilde{X})+h^{2,1}(\widetilde{X})+h^1(End_0(\widetilde{V}))=4+53+298=355.
\end{eqnarray}
From the point of view of the massless heterotic spectrum, it is clear that in the theories associated to the TSD geometries, $(X,V)$ and $(\tilde{X},\tilde{V})$, all the degree of freedom appear to match.

Moreover, we have chosen this pair of geometries to have a further special property. Each CY 3-fold appearing in the dual pair exhibits an elliptic fibration structure. As a result, by the arguments laid out in Section \ref{Background}, we expect each heterotic background in the pair to lead to its own F-theory dual.

A closer inspection yields the following elliptic fibration structures:
\beq
\pi_h: X \stackrel{\mathbb{E}}{\longrightarrow} \mathbb{P}^2~~\text{and}~~~ \tilde{\pi}_h: \tilde{X}  \stackrel{\mathbb{E}}{\longrightarrow} dP_1
\eeq
 The fibrations of $X$ and $\tilde X$ can be seen very explicitly from the form of the complete intersection descriptions of the manifolds (so-called ``obvious" fibrations \cite{Anderson:2017aux}). Below we use dotted lines to separate the ``base" and ``fiber" of the manifold: 
\begin{equation}\label{4d_dual_cy}
X =  \left[ \begin{array}{c|cc}  \IP^1 & 1&1\\ \IP^2 & 1&2\\ \hdashline  \IP^2 &1&2\end{array} \right], \quad\quad
\widetilde X =  \left[ \begin{array}{c|c:cc}  \IP^1 & 0&1&1\\ \IP^2 & 0&1&2\\  \hdashline  \IP^2 & 1 &1&1\\ \IP^1 & 1&0&1\end{array} \right],
\end{equation}
where the base for the elliptically fibered $X$ is $B_2 = \mathbb P^2$ (the bottom row of the configuration matrix) while the $dP_1$ base for $\tilde X$ is given as $\widetilde{B_2} = \left[ \begin{array}{c|c} \IP^2 & 1\\ \IP^1 & 1\\ \end{array} \right]$.

Employing the techniques of \cite{Anderson:2016ler, Anderson:2016cdu}, we find that the fibrations in both $X$ and $\tilde X$ in fact admit rational sections and as a result are elliptically fibered (as opposed to genus-one fibered only). Moreover, each fibration contains two rational sections (i.e. a higher rank Mordell-Weil group). In an abuse of notation, we will use $\sigma_i$ to denote both the two sections to the elliptic fibration of $X$ (respectively $\tilde X$) and the associated K\"ahler forms dual to the divisors.  In terms of the basis of  the K\"ahler  $(1, 1)$-forms $J_r$ inherited from the ambient space factors $\IP^n_r$ of each CICY 3-fold:
\begin{eqnarray}\nonumber
\sigma_1(X) = -J_1 + J_2 + J_3, &&\quad \sigma_2(X) = 2 J_1 - J_2 + 5 J_3 \\
\sigma_1(\widetilde X) = -J_1 + J_2 + J_3, &&\quad \sigma_2(\widetilde X) = 2 J_1 - J_2 + 4 J_3 + J_4,
\end{eqnarray}
With a choice of zero section for each manifold from the set above, the CY 3-fold can in principle be put into Weierstrass form \cite{Deligne, Nakayama}. For explicit techniques to carry out this process we refer the reader to \cite{Anderson:2016ler}.

In summary then, we have produced an explicit example of a TSD pair in which both sides are elliptically fibered manifolds, admitting $4$-dimensional, ${\cal N}=1$ F-theory duals in principle. This is an important point of principle, since we have demonstrated that \emph{some} F-theory correspondence should exist for the dual F-theory EFTs. In practice however, it should be noted that explicitly determining the F-theory duals for the geometries given above is difficult. We will begin untangling this process explicitly in Section \ref{Fth}.

For now we close this example by observing an interesting feature of the TSD pair above: Since we began with a deformation of the tangent bundle, the associated $(0,2)$ GLSM admits a $(2,2)$ locus. However, in the TSD geometry the bundle we obtain is no longer manifestly a holomorphic deformation of the tangent bundle on $\tilde{X}$. It remains an open question whether this second theory admits a $(2,2)$ locus in some subtle way. For the moment, we will turn to one further TSD pair in which neither vector bundle is related to the tangent bundle.

\subsection{More general vector bundles}

Here we present a second example in which the same CY manifolds appear, but with different vector bundles. Once again, we start with the GLSM charge matrix determining the pair (X,V) as in (\ref{eq:eg32}) where in this time we have a rank 4  bundle with structure group $SU(4)$:

\begin{equation}
\begin{aligned}
&\begin{array}{|c|c| |c|c|}
\hline
x_i & \Gamma^j & \Lambda^a & p_l \\ \noalign{\hrule height 1pt}\begin{array}{cccccccc}
1& 1& 0& 0& 0& 0& 0& 0 \\
0& 0& 1& 1& 1& 0& 0& 0 \\
0& 0& 0& 0& 0& 1& 1& 1
\end{array}&\begin{array}{cc}
 -1& -1   \\
 -1& -2  \\
 -1& -2
\end{array}&\begin{array}{cccccc}
1& 0& 0& 0& 0& 1 \\
0& 1& 1& 0& 0& 2 \\
0& 0& 0& 1& 1& 1
\end{array}&\begin{array}{cc}
 -1& -1   \\
 -2& -2  \\
 -2& -1
\end{array}\\
\hline
\end{array}\end{aligned}\label{eq:eg32}\end{equation}

In this case, the second Chern class of  $(X,V)$  is different from (\ref{c2tangent}):
\begin{eqnarray}
&&c_2(V) = 2J_1J_2  +J_2^2 +2J_1J_3 +4J_2J_3 +J_3^2,
\end{eqnarray}
However, in this case, $c_2(V) \leq c_2(TX)$ and thus it is expected that this bundle could be embedded into one factor of the $E_8 \times E_8$ heterotic string, where another bundle $V'$ is embedded into the second factor. By completing the geometry in this way, with $c_2(V)+c_2(V')=c_2(TX)$, the anomaly cancellation conditions can be satisfied (alternatively, NS5/M5 branes might be considered).

Following the standard procedure described above, the target space duality data is given by ($\widetilde X, \widetilde V$) with the following charge matrix:

\begin{equation}\begin{aligned}&\begin{array}{|c|c| |c|c|}
\hline
x_i & \Gamma^j & \Lambda^a & p_l \\ \noalign{\hrule height 1pt}\begin{array}{cccccccccc}
1& 1& 0& 0& 0& 0& 0& 0& 0& 0 \\
0& 0& 1& 1& 1& 0& 0& 0& 0& 0 \\
0& 0& 0& 0& 0& 1& 1& 1& 0& 0 \\
0& 0& 0& 0& 0& 0& 0& 0& 1& 1  
\end{array}&\begin{array}{ccc}
 0& -1& -1   \\
 0& -1& -2  \\
-1& -1& -1 \\
-1&  0& -1
\end{array}&\begin{array}{cccccc}
1& 0& 0& 0& 0& 1 \\
0& 1& 1& 0& 0& 2 \\
0& 0& 0& 1& 0& 2 \\
0& 0& 0& 0& 1& 0
\end{array}&\begin{array}{cc}
 -1& -1   \\
 -2& -2  \\
 -2& -1  \\
 -1&  0
\end{array}\\
\hline
\end{array}\end{aligned}\label{eq:eg32dual}\end{equation}
Here the second chern classes of the tangent bundle and the monad vector bundle are respectively:
\begin{eqnarray}\nonumber
&&c_2(TX) = 3J_1J_2  +J_2^2 +2J_1J_3 +3J_2J_3 +J_1J_4 +2J_2J_4 +2J_3J_4, \\
&&c_2(V) = 2J_1J_2  +J_2^2 + J_1J_3 +2J_2J_3 +J_1J_4 +2J_2J_4 +2J_3J_4,
\end{eqnarray}
which could also satisfy the $c_2$ matching condition with the addition of a hidden sector bundle.  

In this background, the bundle structure group of $H=SU(4)$ breaks $E_8$ to $SO(10)$.  As above, the charged matter content can be determined by  the decomposition under $SO(10) \times SU(4)$:
\bea
{\bf 248}_{E_8} \rightarrow  ({\bf 1, 15}) \oplus  ({\bf 10, 6})  \oplus ({\bf \overline{16},  \overline{4}}) \oplus ({\bf 16, 4}) \oplus ({\bf 45,1}).
\eea
The multiplicity of the spectrum is then determined via bundle-valued cohomology as:
\bea
&{\bf 16}'s:   \qquad  h^1(V) = 48,  \qquad   h^1(\widetilde V) = 48, \\
&{\bf \overline{16}}'s:   \qquad  h^1(V^*) = 0,   \qquad   h^1( \widetilde V^*) =  0 ,\\
&{\bf 10}'s:   \qquad  h^1(\wedge^2 V) =  0,  \qquad    h^1( \wedge^2 \widetilde V) =  0.
\eea
\\
Furthermore, the counting of the gauge singlets appearing in this TSD pair match as well:
\begin{eqnarray} \nonumber
&h^{1,1}(X)+h^{2,1}(X)+h^1(End_0(V))=3+60+159=222 ,\\
& h^{1,1}(\widetilde{X})+h^{2,1}(\widetilde{X})+h^1(End_0(\widetilde{V}))=4+53+165=222.
\end{eqnarray}

\vspace{.25in}

With these two examples in hand, it is clear that at least the first question outlined in Section \ref{sec:intro} can be answered in the positive. \emph{Heterotic TSD pairs can indeed be found in which both halves of the dual pair exhibit elliptic fibrations}. However, it is clear that the manifolds in our examples above are not in simple Weierstrass form (and exhibit higher rank Mordell-Weil group) as a result, their F-theory dual geometries may be difficult to determine using standard tools. We review some of these tools in the subsequent Sections before returning to the two examples above in Section \ref{4D_thoughts}.

%\subfile{ftheory_dual}

\section{Inducing a duality in F-theory}\label{Fth}

\subsection{Essential tools for Heterotic/F-theory duality}

In type IIB superstring theory, the axion-dilaton transforms under $SL(2, Z)$ while leaving the action invariant. 
However, it is frequently assumed the string coupling
$g_s$ vanishes and the backreaction from 7-branes is ignored. As a result, many important non-perturbative
aspects of the string compactification which are crucial both conceptually and phenomenologically, are missing.
%Vafa brings up F-theory that has extra two dimensions always acting as a torus.
This is exactly where F-theory arises as a proper description of orientifold IIB theory with $(p, q)$ 7-branes and varying
finite string coupling (i.e. axion-dilaton). The classical $SL(2, Z)$ self-dual symmetry of Type IIB theory acting 
on the axion-dilaton is identified 
as the modular group of a one complex dimensional torus 
$T^2$ and as the complex
structure of a fictitious elliptic curve. In this way, we formally attach an elliptic curve at each point of the type
IIB space time and promote the 10-dimensional IIB theory to auxiliary 12-dimensional F-theory. This structure defines a genus-one or
elliptic fibration. The locus where the fiber degenerates is where the 7-brane is wrapped in the internal CY. F-theory realizes a remarkable synthesis of geometry and field theory in that the structure of the 7-branes/gauge sector, matter content and Yukawa couplings are all encoded in the geometry of the fibration structure and the back-reaction of these branes is taken into account.

There is no description of F-theory as a fundamental theory, but rather, as duals to other theories. A concrete example would be an $8$-dimensional duality \cite{Vafa:1996xn} 
%({i.e.} heterotic theory compactified on $T^2$ related to F-theory on a $K3$ surface), fibered over a shared base manifold $B_{n-1}$ to obtain lower-dimensional dualities. 
 i.e, F-theory conpactified on $K3$ is dual to type IIB on $S^2$ with 24 7-branes turned on, which is also dual to heterotic on $T^2$. The duality between F-theory and heterotic is described further as  F-theory compactified on $K3$ fibered Calabi-Yau $(n+1)$-fold  is dual to $E_8\times E_8$ heterotic string compactified on Calabi-Yau $n$-fold which is elliptically fibered on the same $(n-1)$-fold base: %Here heterotic refers to $E_8\times E_8$, as most works are based on $E_8\times E_8$.
\begin{itemize}
\item Heterotic: $\pi_h: X_n\stackrel{\mathbb E}{\longrightarrow}B_{n-1}$ elliptic fibration
\item F-theory: $\pi_f: Y_{n+1}\stackrel{K3}{\longrightarrow}B_{n-1}$  where $\rho_f: Y_{n+1}\stackrel{\mathbb E}{\longrightarrow}B_n,$ $\sigma_f: B_n\stackrel{\mathbb P^1}{\longrightarrow}B_{n-1}$
\end{itemize}
The paired heterotic/F-theory geometries given above involves both elliptic and $K3$ fibered manifolds. In particular, the F-theory geometry, $Y_{n+1}$ must be compatibly $K3$ and elliptically fibered. The requirement of these two fibration  implies  that $Y_{n+1}$ should also be elliptically fibered over a complex $n$-dimensional base, ${ B}_n$ which is in turn rationally fibered. The existence of a section in any two of the fibrations structures is enough to guarantee the existence of a section in the third fibration ({i.e.} if $\rho_f$ and $\sigma_f$ both admit sections then so does the fibration $\pi_f$). 

With different number of $n$'s, there are theories in different dimensions. Specifically, $n=1,2,3$ will lead to $8D$, $6D$, and $4D$ respectively. When $n=1$, the $(n-1)$-fold base $B_{n-1}$ is a point, when $n=2$ it is a $\mathbb P^1$.  In 4D case, the duality can be written as:
\begin{equation}
\begin{array}{lllll}
&~~~~Y_{4}&\xrightarrow{~~\mathbb{E}~~}&{ B}_3&\\
 &K3~\Big\downarrow&&~\Big\downarrow~\mathbb{P}^1& \\
&~~~~ B_{2} & \xrightarrow{~~ = ~~}  & {B}_{2}& 
\end{array}
\label{nested_fib}
\end{equation}
%\begin{itemize}
%\item Heterotic: $\pi_h: X_3\stackrel{\mathbb E}{\longrightarrow}B_{2}$ elliptic fibration
%\item
 %F-theory: $\pi_f: Y_{4}\stackrel{K3}{\longrightarrow}B_{2}$  where $\rho_f: Y_{4}\stackrel{\mathbb E}{\longrightarrow}B_3,$ $\sigma_f: B_3\stackrel{\mathbb P^1}{\longrightarrow}B_{2}$
%\end{itemize}
By the fibration structure of the CY $4$-fold \eref{nested_fib}, the base ${ B}_3$ must be $\mathbb{P}^1$-fibered. 
%As in the case of the Hirzebruch surfaces in the $6$-dimensional theory, 
%The simplest class of geometries will correspond to bases that are $\mathbb{P}^1$ bundles over $B_2$. 
As in \cite{Friedman:1997yq}, such a $\IP^1$ bundle can be defined as the projectivization of two line bundles,
\beq\label{p1bund}
{ B}_3=\mathbb{P}(\cO \oplus {\cal L}) \ , 
\eeq
where $\cO$ is the trivial bundle and ${\cal L}$ is a general line bundle on $B_2$. 
In this case the topology of ${ B}_3$ is completely fixed by the choice of line bundle ${\cal L}$ and we can defined  a $(1,1)$-form on $B_2$ as $T= c_1({\cal L})$.  A special case would be $R=c_1(\mathcal O(1))$ where $\cO(1)$ is a bundle that restricts to the usual $\cO(1)$ on each $\mathbb P^1$ fiber.
 They satisfy the relation $R(R+T)=0$ in cohomology class, which indicates the two corresponding sections don't intersect to each other. 
This kind of twist allows us to matching the degrees of freedom in the $4$-dimensional heterotic/F-theory dual pairs.

In the  $E_8\times E_8$ heterotic side, the vector bundle can be decomposed as $V=V_1\oplus V_2$, and the curvature splits as 
\bea
\label{eq:c_2}
c_2(V) = \frac {1}{30} \text{Tr}F_i^2 = \eta_i \wedge \sigma + \xi_i,
\eea
 where $\eta_i, \xi_i$ are pullback of 2-forms and 4-forms on $B_2$, $\sigma$ is the Poincare dual to the section of the elliptic fibration $\pi_h: X_3 \stackrel{\mathbb{E}}{\longrightarrow} B_2$.  For any CY $3$-fold in Weierstrass form as described above, $c_2(TX_3)=12c_1(B_2)\wedge \omega_{\hat 0} +(c_2(B_2) +11c_1(B_2)^2)$ \cite{Friedman:1997yq}. The heterotic Bianchi identity requires $\eta_1 + \eta_2 = 12c_1(B_2)$, which enable us to parameterize a solution as 
 \bea
 \label{eq:twist}
 \eta_{1,2} = 6c_1(B_2)\pm T' ,
 \eea
 where $T'$ is a (1,1) form on $B_2$. 
By studying the 4D effective theories of these dual heterotic/F-theory
compactifications it is straightforward to determine that the defining {(1, 1)}-forms
$T$, $T'$ are identical to each other $T=T'$. 
Then the {(1, 1)}-form $T$ is referred to
as the so-called \lq\lq twist" of the $\IP^1$-fibration and is the crucial defining data of the simplest
classes of heterotic/F-theory dual pairs.
Moreover, 
%the spectral cover construction that maps the vector bundle to an n-sheet cover gives further correspondence.
this  duality map dependences on a
particular method of constructing Mumford poly-stable vector bundles, the spectral cover construction.

\subsection{Spectral Cover Construction}
To find the dual F-theory model of a specific Heterotic model, we need a description of the moduli space of stable degree zero vector bundles over elliptically fibered manifolds, (the standard formulation works for Weierstrass fibration, but it can be generalized to other types of elliptic fibrations) in terms of two ``pieces", which are called spectral data. This can be done by Fourier-Mukai transform, and we briefly review it in the following (Fourier-Mukai is an important type of functor between derived categories, but to avoid unnecessary technicalities, we only restrict ourselves to the following special type defined by Poincare bundle, and ignore general discussions. Interested reader can refer to \cite{HuyFM, BBRH}).\footnote{Please note that we restrict ourselves to $SU(N)$ (degree zero, and stable) vector bundles over a Weierstrass elliptic fibration.}\\

Fourier-Mukai transform takes the stable bundles of degree zero and rank $n$ over $X$, and gives a torsion sheaf (rank $0$) degree $n$ over the (compactified) Jacobian fibration $\tilde{X} \sim X$. More precisely, consider the following fiber product, and the natural projections, 

\begin{displaymath}
    \xymatrix{
           & X\times_B \tilde{X} \ar[dl]_{\pi_1}\ar[d]^{\rho} \ar[dr]^{\pi_2}  &     \\
        X &         B                  & \tilde{X}}
\end{displaymath}

then FM transform is defined as \footnote{In some sense it is similar to Fourier transform. In that case one starts with a function $f(x)$ defined in a space $X$, then we pull back the function into a larger space $X\times Y$, multiply by a kernel $e^{ix \cdot y}$ and integrate over $x$ to get a function in $Y$. The pushforward action is similar to the integration over the fibers}, 

\begin{equation}\label{FMdef}
FM^1 (V) =\mathbf{R}^1{\pi_2}_* ({\pi_1}^* V \otimes \mathcal{P}^{*}),
\end{equation}

We emphasize again(compactifed) Jacobian of irreducible elliptic curves are isomorphic to the elliptic curve, therefore in the Weierstrass fibration we have $X \sim \tilde{X}$. In (\ref{FMdef}) $\mathcal{P}$ is the Poincare bundle,
\begin{equation}
\mathcal{P}= \mathcal{O}(\Delta-\sigma_1 \times \tilde{X} - X \times \sigma_2)\otimes \rho^*\mathcal{K}_B,
\end{equation}
where $\Delta$ is the diagonal divisor in $X\times_B \tilde{X}$, $\sigma_1$ and $\sigma_2$ are section of the first and second factor respectively. $R^1 \pi_*$ is the first derived pushforward. Roughly speaking, the presheaf corresponding to $R\pi_* \mathcal{F}$ over an open set $U$ is isomorphic to $H^1(\pi^{-1}U,\mathcal{F})$ (look at \cite{Hart} III.1 and \cite{Weibel}). Note that the restriction of stable and degree zero vector bundles over generic fibers will be a semistable bundle with degree zero and the same rank \cite{friedmanBook}. Then, a well known theorem proved by Atiyah\footnote{Intuitively, one can argue that to construct a non trivial flat bundle over a torus the only way is to embed the fundumental group of the torus inside the Lie group, i.e. turn on the Wilson lines \cite{FMW}.}\cite{AtiyahElliptic} tells us the the general form of restriction of the vector bundle over a generic smooth elliptic curve $E$ \footnote{Note that $\sum_i rank(\cal E_i) = rank(V)$},

\begin{align}\label{Atiyah}
&V|_E = \bigoplus_i \mathcal{E}_i \otimes \mathcal{O}_E(p_i - \sigma) , \\
&\sum_i rank(\mathcal{E}_i)\times p_i = 0, \nonumber
\end{align}

The second condition is imposed because of the \lq\lq S" of the $SU(N)$ gauge group (note that the sum is over the group law of the elliptic curve), and $\mathcal{E}_i$ is constructed inductively by extending with trivial bundle,
\begin{equation}
0\rightarrow \mathcal{O}_E \rightarrow \mathcal{E}_i \rightarrow \mathcal{E}_{i-1}\rightarrow 0.
\end{equation}
Therefore using (\ref{Atiyah}), it's easy to show that the stalk of the Fourier-Mukai transform of $V$ over some generic point $p\in \tilde{X}$ takes the following form (the isomorphism can be proved by a the final theorem in \cite{Hart} III.12),

\begin{equation}\label{fiber}
FM^1(V)_p \sim H^1 (E_p, \bigoplus_i \mathcal{E}_i \otimes\mathcal{O}_E(p_i-p))
\end{equation} 

It's clear that (\ref{fiber}) is non-zero only at point $p$ that are coincident with one of the points $p_i$. To see this consider the $\mathcal{E}_2$. By definition we have
\begin{equation}
0\rightarrow \mathcal{O}_E \rightarrow \mathcal{E}_2 \rightarrow \mathcal{O}_E\rightarrow 0.
\end{equation}
\noindent
If we multiply the whole sequence with $\mathcal{O}_E(p_i-p)$, the first cohomology of $\mathcal{E}_2 \otimes\mathcal{O}(p_i-p)$ will be zero, since $\mathcal{O}(p_i-p)$ is a line bundle over the elliptic curve corresponding to a degree zero, but non effective divisor (so it doesn't have global section, and by Riemann-Roch, trivial first cohomology). By induction we can get similar conclusions for a general $\mathcal{E}_i$ in (\ref{Atiyah})\\

 Therefore the Fourier-Mukai sheaf $FM^1(V)$ is supported on a n-sheeted cover, $\mathcal{S}$, of the base  (possibly a non-reduced and/or reducible scheme), which is called a spectral cover. To be more clear consider a special case of (\ref{Atiyah}),

\begin{equation}\nonumber
V|_E=\bigoplus_{i=1}^n \mathcal{O}(p_i-\sigma) \\
\end{equation}
such that the points $p_i$ are all different. In fact every semistable bundle over $E$ is S-equivalent to the direct sum written above. In this situation $\mathcal{S}$ is a non degenerate surface (i.e. non reduced as a scheme), and by (\ref{fiber}) rank of the Fourier-Mukai sheaf when restricted over $\mathcal{S}$ is one. Therefore the Fourier-Mukai transform of the vector bundle is described by an n-sheeted cover of the base $\mathcal{S}$, and a sheaf (called spectral sheaf) over that, which if $\mathcal{S}$ is non degenerate, the restriction of the spectral sheaf will be a line bundle $\mathcal{L}$ over $\mathcal{S}$.\footnote{More precisely the rank of the spectral sheaf over the modified support (a scheme defined by 0th Fitting ideal \cite{BBRH}) is one. For example is the support is defined as $z^2=0$, rank of the spectral sheaf over the topological support, $z=0$ locus, is $2$, but over the modified support, which roughly looks two copies of $z=0$ infinitesimally close to each other, the rank of the sheaf is one}  \\

The next question is how the vector bundle can be reconstructed from the spectral data described above. This can be done by using the inverse ``functor", which simply is,
\begin{equation}
V={\pi_1}_* ({\pi_2}^* FM^1(V)\otimes \mathcal{P})
\end{equation}

The final point is that, technically, Fourier-Mukai is the equivalence of the derived category of coherent sheaves on $X$ and derived category coherent sheaves on $\tilde{X}$ (at least when $X$ and $\tilde{X}$ are smooth). It roughly means that, Fourier-Mukai transform gives a ``one to one" relation between the ``space of stable vector bundles $V$" and the ``space of spectral data $(\mathcal{S},\mathcal{L})$" . The logic is that the latter seems easier to study rather than the original bundles.  \\

In principle, it is possible to find the F-theory dual by using the spectral data. We review this briefly in the following. Suppose we have two vector bundles ($V_1 , V_2 $) over a Weierstrass elliptically fibered manifold $X$ (suppose there are no NS5 branes), then heterotic anomaly cancellation requires,

\begin{equation}\nonumber
c_2(V_1)+c_2(V_2)=c_2(X).
\end{equation}
Then second Chern classes (which can be computed by Grothendieck-Riemann-Roch, if we have the spectral data \cite{FMW}), can be written generally as,

\begin{align}\label{twist}
&c_2(V_i) = \sigma \eta_i +\omega_i,  \nonumber \\
&\eta_i=6 C_1(B_H) \pm T,
\end{align}
where $\eta_i$ is a divisor in the base ($B_H$), and $\omega_i$ is the intersection of two divisor in $B_H$. Also by using the same method it is not too hard to show that the divisor class of the spectral cover of $V_i$ is given by 
\begin{eqnarray}
[\mathcal{S}] = n_i \sigma+\eta_i. 
\end{eqnarray}
Now, the first statement about the Heterotic and F-theory duality is that the topology of the base manifold of the F-theory Calabi-Yau is fixed by the ``twist" $T$ in  (\ref{twist}) as,

\begin{equation}\label{Fbase}
B_F= \mathbb{P} (\mathcal{O}_{B_H}\oplus \mathcal{O}_{B_H}(T))
\end{equation}
The second statement is that the complex structure of $\mathcal{S}$, (partially) fix the complex structure of the Calabi-Yau in the F-theory side. It's easier to describe this with an example. Suppose we have a $SU(2)$ bundle $V$, and it's spectral cover is non degenerate,

\begin{equation}
S=a_0 z^2 +a_2 x.
\end{equation}
  Since one of the $E_8$ factors breaks to $E_7$, we should have an $E_7$ singularity in the F-theory geometry, which is described by the following Weierstrass equation,

\begin{align}
&Y^2=X^3+F(u,z) x+ G(u,z), \nonumber\\
&F=\Sigma_{i=1}^8 F_i (z) u^i, \nonumber\\
&G=\Sigma_{i=1}^{12} G_i(z) u^i.
\end{align}
where $u$ is the affine coordinate of the $\mathbb{P}^1$ fiber of (\ref{Fbase}), and $z$ is the ``collective" coordinate for $B_H$. Now, the conjectured duality tells us the corresponding $E_7$ singularity should be located near $u=0$, therefore $F_0=F_1=F_2=0$, and $G_0=\dots=G_4=0$. Also $a_0$ is identified with $G_5$, and $a_2$ with $F_3$. \\
The other vector bundle (which is embedded in the other $E_8$ factor) determines the singularity near $u\rightarrow \infty$, and higher polynomials ($F_5, \dots$ and $G_7 \dots$) are determined by the spectral cover of that vector bundle (which we didn't write here). The middle polynomials $F_4$ and $G_6$ are determined by the Heterotic Weierstrass equation.\\

The last piece of data which is remained is the spectral sheaf $\mathcal{L}$, which is an element pf the Picard group $Pic(S)$. The ``space of line bundles" itself is made by two pieces, the ``discrete" part $H^{1,1}(S)$, and the ``continuous" part which is $J(S)$ (space of degree zero (flat) line bundles),
\begin{equation}\label{Picard}
0\rightarrow J(S)\rightarrow Pic(S) \rightarrow H_{\mathbb{Z}}^{1,1}(S)\rightarrow 0.
\end{equation}   
  In 6D theories, the discrete part can be fixed uniquely by using Fourier-Mukai transform, and the Jacobian of the curve is mapped to the intermediate Jacobian of the Calabi-Yau threefold in F-theory. In type IIA or M-theory language, the ``space of three forms", $H^3(X,\mathbb{R})/H^3(X,\mathbb{Z})$, is described by intermediate Jacobian \cite{FMW, Aspinwall, Curio:1998bva}.

The situation in 4D theories are even more complicated. In such cases, it is possible to have non trivial 4-form fluxes which can be introduced in various (equivalent) ways. One way is to define is as 4-form induced by the field strength of the 3-form in M-theory limit. Another way is to define as a (1,1)-form flux over the 7-branes wrapping the divisors in the base times another (1,1)-form localized around the 7-brane locus \cite{Denef:2008wq, Donagi:2008ca}. In general, the 4-flux data is parameterized by the Deligne cohomology (See the lectures \cite{Weigand:2018rez}, and references there.)

\begin{equation}\label{G4}
0\rightarrow J^2(\hat{X}_4)\rightarrow H^4_D(\hat{X}_4,\mathbb{Z}(2)) \rightarrow H_{\mathbb{Z}}^{2,2}(\hat{X}_4)\rightarrow 0,
\end{equation} 
where $\hat{X}_4$ is the resolved geometry in M-theory limit, $J^2$ is the intermediate Jacobian, 
\begin{eqnarray}
J^2(\hat{X}_4) = H^3(\hat{X}_4,\mathbb{C})/(H^{3,0}(\hat{X})\oplus H^{2,1}(\hat{X}_4)),
\end{eqnarray}
which corresponds to the space of flat 3-forms in M theory .The third, and most difficult part, of the Heterotic F-theory duality, is that continuous part of the spectral sheaf data, $J(S)$ maps to $J^2(X_4)$, and discrete part, $H^{1,1}(S)$ (which is determined by the divisor class (first Chern class) of the spectral line bundle), maps to the discrete part of the 4-flux data $H^{2,2}(\hat{X}_4)$.

%\subfile{eg_6D}

%%%%%%%%%%%%%%%%%%%%%%%%%%%%%%%%%%%%%

\section{Warm up: Heterotic/F-theory duality in $6$-dimensions}\label{6D_Section}

%%%%%%%%%%%%%%%%%%%%%%%%%%%%%%%%%%%%%
In this section, we'll begin in earnest the process of attempting to determine the induced duality in F-theory given by TSD and whether the multiple fibrations conjecture outlined in previous sections could be a viable realization. In this simpler context both the geometry of the F-theory compactification as well as the process of reparameterizing (i.e. performing a FM-transform) of the heterotic data are more readily accomplished.

To begin, it should be noted that in the context of heterotic target space duals we will consider smooth geometries (i.e. smooth bundles over $K3$ manifolds) in the large volume, perturbative limit of the theory. We will consider solutions without NS 5-branes so that the 6-dimensional theory exhibits a single tensor multiple (associated to the heterotic dilaton) (see \cite{Taylor:2011wt} for a review). Within the context of 6-dimensional F-theory EFTs with a single tensor and a heterotic dual it is clear that we must restrict ourselves to CY 3-folds that are elliptically fibered over Hirzebruch surfaces:
\beq
\pi_f: Y_3 \to \mathbb{F}_n
\eeq
with $n \leq 12$ \cite{Morrison:1996na,Morrison:1996pp}. 

It is our goal in this section to test the multiple fibrations conjecture in the context of target space duality. At the level of GLSMs TSD in heterotic compactifications on $K3$ works mechanically exactly as in the case of CY 3-folds. However, the associated geometry is dramatically simpler. It is clear that the two TSD GLSMS will parameterize at best two different descriptions of a $K3$ surface and that the process must by construction preserve the second chern class of the vector bundle $V$ over $K3$ (see \cite{Blumenhagen:2011sq} for a proof valid for either CY 2- or 3-folds). Since the massless spectrum of the $6$-dimensional heterotic theory compactified on a smooth $K3$ is entirely determined by the rank and second Chern class of $V$ (see e.g. \cite{Bershadsky:1996nh})\footnote{And the moduli space of stable sheaves over $K3$ with fixed Chern character has only one component.}, it is clear that TSD is only a simple  re-writing of the same geometry and $6$-dimensional EFT.

However, there remains something interesting to compare to in that it can still be asked: \emph{Does the concrete process of heterotic TSD duality in $6$-dimensions correspond to exchanging $K3$ fibrations in the dual F-theory geometry?} In the context of F-theory 3-folds that are elliptically fibered over a Hirzebruch surface as described above, there is in fact only one geometry where multiple $K3$ fibrations can arise. It was established in the very first papers on F-theory \cite{Morrison:1996na,Morrison:1996pp}, that in order to have different $K3$ fibrations within a CY threefold with a perturbative heterotic dual, the base twofold must be $F_0$. Indeed, the remarkable observation of Morrison and Vafa was that the existence of two $K3$ fibrations in $\pi_f: Y_3 \to \mathbb{F}_0$ (only a simple relabeling in F-theory) was dual to a highly non-perturbative heterotic/heterotic duality discovered by Duff, Minasian and Witten \cite{Duff:1996rs}.

With these observations in mind, we can immediately make several observations. To begin, we must recall that a base manifold for the F-theory fibration of $\mathbb{F}_n$ in this context is correlated to bundles with $c_2=12 \pm n$ in the $E_8 \times E_8$ heterotic dual. Thus
\begin{itemize}
\item For any purely perturbative heterotic TSD pair in $6$-dimensions with $c_2(V)=c_2(\tilde{V})\neq 12$, $(0,2)$ TSD \emph{cannot correspond to multiple fibrations in F-theory} (since as described above, such multiple $K3$ fibrations arise only for $n=0$).
\item With the point above, we have established that in general the multiple fibrations conjecture outlined in introduction the \emph{must be false in general} at least in the 6-dimensional heterotic theories. 
\item This demonstrates that not all TSD pairs can be described by F-theory multiple fibrations, but the converse question, namely -- \emph{can multiple fibrations in F-theory give rise to dual heterotic TSD pairs?} -- in principle remains open.
\end{itemize}
Thus, in this section we chose to look at this last point in closer detail by considering an example TSD pair over $K3$ in which $c_2(V)=c_2(\tilde{V})=12$ (the so-called symmetric embedding), corresponding to an $\mathbb{F}_n=\mathbb{P}^1 \times \mathbb{P}^1$ base in F-theory. This will at least make it possible in principle for the two duals to consider.\\

\subsection{Spectral Cover of monads}

To begin, we observe that since the vector bundles defined by GLSMs (in their geometric phases) are usually presented as monads \cite{horrocks}, we must deal with how to convert this description of a bundle into one compatible with heterotic/F-theory duality. As discussed in Section \ref{Fth}, it is necessary to perform an FM transform to compute the spectral cover is this case. As a result, here we briefly review the method first introduced in \cite{Ber}. \\

Suppose we are given a general monad such as,
\begin{equation}\label{monadBer}
0\rightarrow \mathcal{V}\rightarrow \mathcal{H}\overset{F}\rightarrow\mathcal{N}\rightarrow 0,
\end{equation}
where $\mathcal{H}$ and $\mathcal{F}$ are direct sum of line bundles of appropriate degrees. If we assume $\mathcal{V}$ is stable and degree zero, then from the previous subsection we know that its restriction over a generic elliptic fiber will look like $\oplus_i \mathcal{O}(p_i-\sigma)$. So if we twist the whole monad by $\mathcal{O}(\sigma)$, 
\begin{equation}
\mathcal{\tilde{V}}:=\mathcal{V}\otimes\mathcal{O}(\sigma) |_E=\bigoplus_i \mathcal{O}(p_i).
\end{equation}
Each factor has only one global section over the fiber, and it becomes zero exactly at the point $p_i$ which is the intersection of the spectral cover with the fiber. So the idea is try to find the global section of the twisted vector bundle over elliptic fibers, then check at what points the dimension of the vector space generated by global sections drop. To illustrate how this can be done explicitly, first consider twisting the full monad sequence by $\mathcal{O}(\sigma)$, 
\begin{equation}\label{monadBertwist}
0\rightarrow \mathcal{\tilde{V}}\rightarrow\tilde{\mathcal{H}}\overset{F}\rightarrow\tilde{\mathcal{N}}\rightarrow 0
\end{equation}
and follow this by next taking the action of the left exact functor $\pi_*$ on the above sequence ($\bar{F}$ is the induced map, corresponding to $F$):
\begin{equation}
0\rightarrow\pi_* \mathcal{\tilde{V}}\rightarrow\pi_* \tilde{\mathcal{H}}\overset{\bar{F}}\rightarrow \pi_*\tilde{\mathcal{N}}\rightarrow R^1\pi_* \mathcal{\tilde{V}} \rightarrow \dots
\end{equation}

If we assume the vector bundle is semistable over every elliptic fiber (this need not be true always, as we'll see. It is only necessary that vector bundle be semistable over generic elliptic fibers), then $R^1\pi_* \mathcal{\tilde{V}}$ is identically zero, because its presheaf is locally of the form $H^1(E,\mathcal{O}(p_i))$. Now consider the action of the right exact functor $\pi^*$ over the last sequence, since the elliptic fibration map $\pi$ is a flat morphism, it doesn't have higher left derived functors (we have $Tor_{1S} (M, R)=0$ due to the flatness, where $S$  is the ring that corresponds to $\mathcal{O}_B$, $R$ corresponds to $\mathcal{O}_X$, and $M$ is the free module corresponding to $\mathcal{V}$ (see e.g. \cite{Weibel}, Chapter 3) ). So we get,

\begin{equation}\label{monadBer2}
0\rightarrow\pi^*\pi_* \mathcal{\tilde{V}}\rightarrow\pi^*\pi_* \tilde{\mathcal{H}}\overset{\bar{F}}\rightarrow \pi^*\pi_*\tilde{\mathcal{N}}\rightarrow 0
\end{equation}

Note, $\pi^*\pi_* \mathcal{\tilde{V}}$ is vector bundle that its fibers over a point $p$ are generated by the global sections of $\tilde{\mathcal{V}}$ over elliptic curve $E_b$, where $b=\pi(p)$. So \eref{monadBer2} tells us that if we find the global sections of $\tilde{\mathcal{H}}$ and $\tilde{\mathcal{N}}$, then the kernel of the induced map $F$ is isomorphic to $\pi^*\pi_* \mathcal{\tilde{V}}$. So the locus where the rank of the kernel drop, coincides with the spectral cover. To clarify these rather abstract ideas, in the following subsection, we explicitly compute the spectral cover of two examples which will be used in the final subsection.

\subsection{Examples} 

To begin we assume that the $K3$ can be written in the following simple toric form,

\begin{equation}\label{K31}
\begin{aligned}&\begin{array}{|c|c|}
\hline
x_i & \Gamma^j \\ \noalign{\hrule height 1pt}\begin{array}{ccccc}
3& 2& 1& 0& 0  \\
6& 4& 0& 1& 1  
\end{array}&\begin{array}{c}
 -6 \\
 -12 
\end{array}\\
\hline
\end{array}
\end{aligned}
\end{equation}
\noindent

$\bullet$ \textbf{Example 1} The first example is the following $SU(2)$ monad,
\begin{equation}\label{GLSM1}
\begin{aligned}&\begin{array}{|c|c|}
\hline
\Lambda & p \\ \noalign{\hrule height 1pt}\begin{array}{cccc}
1& 1& 2& 3  \\
1& 5& 3& 7  
\end{array}&\begin{array}{cc}
 -3& -4 \\
 -9& -7 
\end{array}\\
\hline
\end{array}
\end{aligned}
\end{equation}

The second Chern class of this monad is $12$. The map $F$ of the monad is given by the following generic matrix,
\begin{equation}
F \sim \begin{bmatrix}\label{bmatrix}
x f_4 +z^2 f_8 & a x+ z^2 g_4 & z f_6 & f_2\\
b y +x z g_2 + z^3 g_6 & z^3 h_2 & c x+ z^2 h_4 & d z
\end{bmatrix}
\end{equation}
where subscripts indicate the degree of homogeneous polynomials over $\mathbb{P}^1$.
With this choice, it can be verified that the kernel of $\bar{F}$ in \eref{monadBer2} takes the following generic form,
\begin{equation}\label{ker1}
\begin{bmatrix}
x-z^2 \frac{-f_2 h_4+d f_6}{c f_2} & 0\\
0 & x\frac{c f_2}{f_2 h_4 -d f_6} +z^2 \\
-\frac{b}{c} y -x z \frac{f_2 g_2-b f_4}{c f_2} - z^3\frac{f_2 g_6-d f_8}{c f_2}& -z^3\frac{f_2 h_2-d g_4 }{f_2 h_4-d f_6} + -x z \frac{a d}{d f_6-f_2 h_4} \\
- x^2\frac{f_4}{f_2}+ y z \frac{b f_6}{c f_2}-x z^2\frac{c f_8 +f_4 h_4-g_2 f_6}{c f_2} - z^4 \frac{f_8 h_4- f_6 g_6}{c f_2}  &  x^2 \frac{a c}{d f_6-f_2 h_4}-x z^2 \frac{c g_4+a h_4}{f_2 h_4-d f_6}-z^4\frac{-f_6 h_2+ g_4 h_4}{f_2 h_4-d f_6}
\end{bmatrix}
\end{equation}
\noindent
where $f_i,g_i,h_i$ are polynomials in terms of base coordinates with degree $i$, and $a,b,c,d$ are constant. The common factor of the minors of \eref{ker1} is,
\begin{equation}\label{incSpec1}
c f_2 x + (f_2 h_4-d f_6) z^2
\end{equation}
\noindent
From the previous discussion naively we might conclude that \eref{incSpec1} must be the spectral cover. However, in the Fourier-Mukai discussion it was noted that the divisor class of the spectral cover should be $2\sigma+12 D$. So, in the expression above we are clearly missing a degree 6 polynomial in \eref{incSpec1}. The correct spectral cover should be
\begin{equation}\label{sp1}
S=F_6 (c f_2 x + (f_2 h_4-d f_6) z^2)
\end{equation}
\noindent
But why then is $F_6$ is missing? The reason is that in the previous subsection we assumed the vector bundles are semistable over every fiber. This not necessarily true. It is possible to start from a stable bundle, and modify it in a way that it becomes semistable over every fiber \cite{FMW2}. \\

To see clearly what happens, let us first find the elliptic fibers such that vector bundle over them is unstable. Note that from \eref{incSpec1} it can be seen that the spectral cover is a non degenerate two sheeted surface, and over generic $E$, $V|_E=\mathcal{O}(p-\sigma)\oplus \mathcal{O}(q-\sigma)$, where $p+q=2\sigma$. So $V|_E$ does not have global section over almost every fiber except when $p=q=\sigma$. These points are on the intersection of $z=0$ and the spectral cover, which are the zeros of $f_2 F_6$. The idea then is to see if we can find the elliptic fibers which over the vector bundle (not its twisted descendant) can have global section. So all we need to do to find $F_6$ is to study the kernel of the induced map in the following sequence:

\begin{equation}\label{missing1}
0\rightarrow\pi_* \mathcal{V}\rightarrow\pi_* \mathcal{H}\overset{F_{ind}}\rightarrow \pi_*\mathcal{N}\rightarrow R^1\pi_* \mathcal{V} \rightarrow \dots
\end{equation}
\noindent
Where the induced map, $F_{ind}$, in the above case is a $7\times 7$ matrix in terms of the base coordinates. Generically it's rank is 7, excepts over the zeros of $f_2 F_6$, so we can read the missing polynomial from this form. Please note that the above computations are local, globally $\pi_* \mathcal{V}=0$. Because $\mathcal{V}$ is locally free, and $\pi$ is a flat morphism, so the pushforward of $\mathcal{V}$ should also be torsion-free (see \cite{Hart}, Ch. III.9.2).

Interestingly, when we repeat the same analysis after twisting with $\mathcal{O}(\sigma)$ and $\mathcal{O}(2\sigma)$, the rank of the corresponding induced map drops over $F_6$, and nowhere respectively. This means that the bundle over those fibers takes the following form,
\begin{equation}
\mathcal{V}|_{E_{\text{missing}}}=\mathcal{O}(p)\oplus \mathcal{O}(-p),
\end{equation}
therefore $h^0(E,\mathcal{V} \otimes \mathcal{O}(\sigma)) \geq 2$, and the rank of the kernel in \eref{monadBer2} (which generically is the same as the rank of the bundle, in this case, 2) doesn't drop over the points of these fibers. So the algorithm suggested in \cite{Ber} doesn't find them. But it can be seen from \eref{FMdef} that these (whole) elliptic fibers are in the support of the spectral sheaf. In summary, a detailed analysis along the lines sketched above shows that the missing component is given by,
\begin{eqnarray}
F_6 &=& h_2c f_2^2-c d f_2 g_4+a d f_2 h_4 - a d^2 f_6.
\end{eqnarray}
where each in the expression polynomial above is defined from the monad map in \eref{bmatrix}.

$\bullet$ \textbf{Example 2} The second example is interesting because its spectral cover is degenerate (in this case, a non reduced scheme):

\begin{equation}\label{GLSM2}
\begin{aligned}&\begin{array}{|c|c|}
\hline
\Lambda & p \\ \noalign{\hrule height 1pt}\begin{array}{cccc}
0& 1& 2& 3\\
2& 1& 3& 7
\end{array}&\begin{array}{cc}
 -2& -4 \\
 -6& -7 
\end{array}\\
\hline
\end{array}\end{aligned}
\end{equation}

The second Chern class of this monad is $c_2(V)=6$, so from the previous discussion, it is clear that the divisor class of its spectral cover must be $2\sigma+6 D$. The number of global sections of $\mathcal{H}$ and $\mathcal{N}$ is seven and six respectively, which tells us that the kernel of $F_{ind}$ is at least one dimensional over almost every elliptic fiber. Since $\mathcal{V}$ is a stable, rank two bundle with $c_1(\mathcal{V})=0$, we conclude either $\mathcal{V}|_E = \mathcal{O}\oplus \mathcal{O}$ or $\mathcal{V}|_E=\epsilon_2$\footnote{By $\mathcal{V}|_E=\epsilon_2$ we mean the unique non trivial extension of the trivial bundles over the elliptic curve.}. In both cases the spectral cover must have the following general form, 
\begin{equation}\label{sp2}
S=F_6 z^2.
\end{equation}

In fact, by computing the $F_{ind}$ directly, we can see the rank of the kernel is always one, so  $\mathcal{V}|_E=\epsilon_2$. As before we can compute the kernel of \eref{monadBer2}, to generate the spectral cover, 
\begin{equation}\label{ker2}
\begin{bmatrix}
z & 0 \\
0 & z^2 \\
-\frac{x z}{f_3}-z^3 \frac{f_4}{f_3} &  z^3 \frac{f_5}{f_3}\\
a \frac{x^2}{f_3}+x z^2 \frac{f_4}{f_3}+z^4\frac{f_8}{f_3} & b y z + x z^2 \frac{g_5}{f_3}+z^4 \frac{f_9}{f_3}
\end{bmatrix}
\end{equation}

Then we look at the minors, and common factor should be spectral cover. However, as in the previous examples, the algorithm in \cite{Ber} miss the polynomial $F_6$. The reason is similar to the previous example, the bundle is unstable over the zeros of $F_6$. It can be shown that the correct spectral cover is indeed,

\begin{equation}\label{sp22}
S=(F_3)^2 z^2,
\end{equation}
where $F_3$ is the entry $(1,3)$ of the monad's map (i.e. the map between the line bundles $\mathcal{O}_X(2,3)$ and $\mathcal{O}_X(2,6)$).\footnote{After a thorough computation we can show that,
\[0\longrightarrow \mathcal{J} \longrightarrow FM^1(V) \longrightarrow \mathcal{O}_{\sigma} \longrightarrow 0,\]
where $\mathcal{J}$ is a torsion sheaf supported over $\left\lbrace z=0 \right\rbrace \cup \left\lbrace F_3=0 \right\rbrace$, and its rank over $\left\lbrace F_3=0 \right\rbrace$ is 2. $\mathcal{J}$ can be computed explicitly, but it is outside of the scope of this paper. 
} 
\subsection{Counter examples of the conjecture}

Here we return to the main goal of this section. Suppose we have two target space dual GLSMs that describe different stable bundles over elliptic $K3$ surfaces. The goal is to check whether their F-theory dual geometries can be related via a change in $K3$- fibrations (i.e. a change in $\mathbb{P}^1$ fibrations in the two-fold base of the CY 3-fold). Generally the base of the F-theory threefold will be a Hirzebruch surface $F_n$, where $n$ is given by the twist in \eref{twist}. The only situation that can accommodate such multiple fibration structures is when $n=0$. So here we focus on this case and demand that the second Chern class of both heterotic vector bundles be $12$.\\

We assume that one of the target space dual geometries is given by monads on the toric $K3$ \eref{K31}.  We also write the elliptically fibered $K3$ in Weierstrass form,
\begin{equation}
y^2=x^3+f_8(x_1,x_2) x z^4+f_{12}(x_1,x_2) z^6,
\end{equation}
where $x_1$ and $x_2$ are the coordinates of the base $\mathbb{P}^1$. To find explicit examples for target space duality, recall there are several constraints that must be met. First, it is necessary to have a well defined GLSM. This means the first Chern class of both bundles should be zero, and the second Chern class of both bundles (or sheaves) should be $12$.

In addition we must make sure that the hybrid phase in which we do the TSD ``exchange" of $G$ and $F$ actually exists. In the process of generating the TSD pairs, it may happen that singularities arise the in bundle or manifold (we expect that crepant resolutions should exist for the manifold and that the singularities in the ``bundle" should be codimension 2 in the base, so that the sheaf is torsion free). In addition to these constraints for the GLSM, there is another practical requirement for finding the F-theory geometry, we prefer to work with $SU(N)$ bundles which have a non-degenerate spectral cover. If this is not satisfied it is still possible to find the F-theory dual, but we should remember that the form of the spectral sheaves can be vastly more rich/complex in these cases. This enhanced data in the Picard group will not be manifest in the spectral cover, or in the complex structure moduli of the dual F-theory geometry. Instead it will be related in the dual F-theory to nilpotent Higgs bundles over singular curves \cite{Aspinwall:1998he,Donagi:2011jy,Donagi:2011dv,Anderson:2013rka}. 

It is straightforward to find many GLSMs where at least one of the bundles (say $\mathcal{V}_1$) has $SU(N)$ structure, and spectral cover is non degenerate, for example consider \eref{GLSM1} once again,

\begin{equation}
\begin{aligned}&\begin{array}{|c|c|}
\hline
\Lambda & p \\ \noalign{\hrule height 1pt}\begin{array}{cccc}
1& 1& 2& 3  \\
1& 5& 3& 7  
\end{array}&\begin{array}{cc}
 -3& -4 \\
 -9& -7 
\end{array}\\
\hline
\end{array}\end{aligned}\nonumber
\end{equation}
with Chern class
\begin{equation}
C_2(V_1)=5\sigma^2+22\sigma D+23 D^2=12~.
\end{equation} 

It should be noted that the algorithm for determining the spectral cover, using the methods of \cite{Ber} was sketched above, but when the spectral cover becomes reducible (which can still be reduced), it is not guaranteed that those methods find the full spectral cover (i.e. usually some (vertical) components will be missed). One can find these components by closer examination of the morphisms that define the bundle and elliptic fibration, as we saw in the last subsection. The spectral cover (schematically) is then given by \eref{sp1}
\begin{equation}
S=F_6 (f_2 x + f_6 z^2)\nonumber
\end{equation}
Note that \eref{GLSM1} by itself is not a well defined linear sigma model, therefore we need another bundle such that it's structure group embedded in the other $E_8$ factor. This second bundle must also have GLSM description over the same $K3$, and it's second Chern class should be,
\begin{equation}\label{Chern2}
C_2(V_2)=6\sigma^2+24 \sigma D +21 D^2=12.
\end{equation}
 Since the existence of this bundle with above properties may not be quite obvious, we turn now to constructing appropriate examples explicitly.

\subsection{Example 1}
We can construct an example $\mathcal{V}_2$ (though not the most general such bundle) as a direct sum of two bundles, each defined by the monad in \eref{GLSM2} (which we denote it here by $\mathcal{V}_0$), with $c_2(\mathcal{V}_0)=6$:
\begin{equation}\label{sum_bun}
 \mathcal{V}_2=\mathcal{V}_0 \oplus \mathcal{V}_0 
\end{equation}
For this monad bundle, the spectral cover was found to be of the form given in \eref{sp22}. In addition, the rank $1$ sheaf on the spectral cover can be readily constrained. Here $ FM^0 (\mathcal{V}_0) $ is zero by results in \cite{Hart} (see Section III.12, the final theorem). So we actually have the following short exact sequence:
\begin{equation}
FM^1(\mathcal{V}_2)=FM^1 (\mathcal{V}_0)  \oplus FM^1(\mathcal{V}_0) 
\end{equation}
where the support of $FM^1(\mathcal{V}_2)$, which is the spectral cover of $\mathcal{V}_2$, is the union of the spectral covers associated to the two copies of $\mathcal{V}_0$. The resulting spectral cover is a non-reduced scheme, which can be realized by the following polynomial\footnote{Generally the two $\mathcal{V}_0$ in the above construction can be related by a continuous deformation, so we consider $F_3$ and $G_3$ as different generic degree 3 polynomials.}
\begin{equation}\label{SPEC2}
S(V_2)=(F_3)^2 (G_3)^2 z^4.
\end{equation} 

Before turning to the F-theory dual of this geometry, let us first construct a target space dual model for the above GLSM. To do that we add new chiral fields, in a way that after integrating them out, we return to the initial model. This can done by adding ``repeated entries" to the charge matrix of the $K3$, and can lead to multiple TSD geometries (all still of the same topological type of manifold and bundle, of course). One possibility is
\begin{equation}\label{Dual GLSM}
\begin{aligned}&\begin{array}{|c|c|c|c|c|c|}
\hline
x& \Gamma & \Lambda_1 & p_1 & \Lambda_2 & p_2 \\ \noalign{\hrule height 1pt}\begin{array}{cccccc}
3& 2& 1& 0& 0& 0\\
3& 2& 0& 1& 1& 1
\end{array}&\begin{array}{cc}
 -6& 0 \\
 -6& -2 
\end{array}&\begin{array}{ccccc}
1& 1& 2& 3& 2\\
0& 4& 1& 5& 6
\end{array}&\begin{array}{ccc}
 -3& -4& -2 \\
 -6& -3& -7 
\end{array}&\begin{array}{cccc}
0& 1& 2& 3\\
2& 0& 1& 4
\end{array}&\begin{array}{cc}
 -2& -4 \\
 -4& -3 
\end{array}
\\
\hline
\end{array}\end{aligned}
\end{equation} 
 
This heterotic geometry ($K3$ manifold and bundle) has point like singularities in the would-be bundle -- that is, it is a rank 2 torsion free sheaf rather that a vector bundle \cite{Distler:1996tj}. \\

With this pair of TSD bundles over $K3$ in hand, we are now in a position to consider the dual F-theory geometry. In this case we will ask the key question: \emph{are the two GLSMs/geometries (i.e. that defined by $\mathcal{V}_1$ and $\mathcal{V}_2$, and its  TSD dual in \eref{Dual GLSM},  realized as different fibrations of a single F-theory geometry?}

By the results of the previous subsection, the complex structure of the Calabi-Yau threefold can be readily determined:
\begin{align}\label{FthCompl}
& Y^2=X^3+F(u_1,u_2,x_1,x_2) X Z^4+G(u_1,u_2,x_1,x_2) z^6,\nonumber\\
&F(u_1,u_2,x_1,x_2) =u_1^4 u_2^4 f_8(x_1,x_2)+u_1^3 u_2^5 F_6(x_1,x_2) f_2(x_1,x_2),\nonumber\\
&G(u_1,u_2,x_1,x_2)=u_1^7 u_2^5 (F_3(x_1,x_2))^2 (G_3(x_1,x_2))^2+u_1^6 u_2^6 f_{12}(x_1,x_2)+u_1^5 u_2^7 F_6(x_1,x_2) f_6(x_1,x_2).
\end{align}
As frequently happens with degenerate spectral data, we find that the apparent F-theory gauge symmetry seems in contradiction with what is expected from the heterotic theory we have engineered. By inspection of the discriminant of \eref{FthCompl}, it is straightforward to see that there appears to be an $E_7$ symmetry on $u_1\rightarrow0$, and an apparent $E_8$ singularity above the curve $u_1\rightarrow \infty$. This might seem in contradiction with the expected gauge symmetry of $SO(12)$ in the hidden sector (determined as the commutant of the $SU(2)\times SU(2)$ structure group defined by the reducible bundle in \eref{sum_bun}). However, in the case of degenerate spectral covers, we naturally expect that T-brane type solutions \cite{Aspinwall:1998he,Anderson:2013rka} may well arise in the dual F-theory geometry. That is, we expect a nilpotent $SU(2)\times SU(2)$ Higgs bundle over the 7-brane which wraps around this curve ($u_2=0$) and breaks the space time gauge group to $SO(12)$ as expected (see \cite{Anderson:2017rpr} for a similar construction).

\begin{figure}[h]
\centering
\includegraphics[scale=0.5]{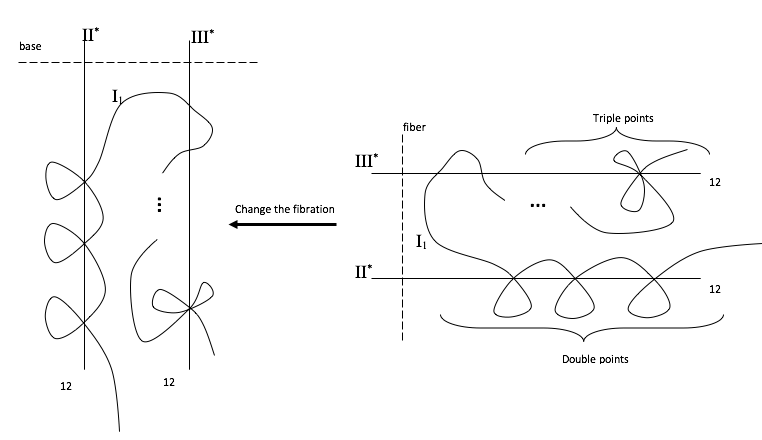}
\caption{The vertical dotted line on the right hand side is the ``vertical $\mathbb{P}^1$". After change of fibration on the left hand side the same $\mathbb{P}^1$ will be the base of the dual heterotic $K3$.}\label{ASP1}
\end{figure}

Next, as demonstrated in \cite{Morrison:1996pp}, changing the $K3$-(resp. $\mathbb{P}^1$-)fibration in the F-theory geometry simply amounts to exchanging the vertical $\mathbb{P}^1$ (whose coordinates are $u_1$ and $u_2$) with the horizontal $\mathbb{P}^1$ which is the base in the initial heterotic $K3$ surface in \eref{K31}. This means that the vertical $\mathbb{P}^1$ becomes the base of a dual heterotic $K3$-surface. To determine gauge groups in the dual heterotic theory, the discriminant curve must be considered. This is shown in Figure \ref{ASP1}. The figure on the right hand side, shows the discriminant of the F-theory Calabi-Yau 3-fold. The line at the top is the locus of the $E_7$ singularity and the one at the bottom corresponds to $E_8$. The curve is the locus of the $I_1$ singularities. It intersects eight times with the $III^*$ curve, where on six of them it has triple point singularities. These six points are exactly the zeros of $F_6$ in Figure \ref{ASP1}, which naively correspond to point like instantons that are responsible for the vertical components in spectral cover of $\mathcal{V}_1$ (not taking into account the spectral sheaf/T-brane data). Similarly the $I_1$ curve intersect with $II^*$ at two sets of three points, which are the zeros of $F_3$ and $G_3$ in \eref{SPEC2}, and over all of them the curve has double point singularities.

We can expand the polynomials in \eref{FthCompl} in terms of $x_1$ and $x_2$, and read the dual heterotic complex structure from there. Clearly we see that the elliptic $K3$ in the new Heterotic dual \emph{must be singular}. In particular, it exhibits singular $E_8$ and $E_7$ located at $u_1=0$ and $u_1\rightarrow \infty$ respectively  (with expected instanton number of $12$ on each locus). This is a highly non-perturbative limit of the string theory. This exchange of gauge symmetry with singularities in the base $K3$ surface arising in the heterotic theory seems to be a generic feature of exchanging F-theory fibrations \cite{Anderson:2016cdu}. As a result, it seems is impossible to get something which is purely smooth/perturbative on both sides like \eref{Dual GLSM} from such a change of fibrations. This shows that at least some of the TSD dual pairs cannot be seen simply as different fibrations of the F-theory geometry. We explore these possibilities a little more in one further example.

\subsection{Example 2}

In this example, the starting geometry/bundles, are the same as before, but here we present another TSD geometry that can also be described easily by spectral cover. So once again we take as our starting point the manifold/bundle:
 
 \begin{equation}\begin{aligned}&\begin{array}{|c|c| |c|c|}
\hline
x_i & \Gamma^j & \Lambda^a & p_l \\ \noalign{\hrule height 1pt}\begin{array}{ccccc}
3& 2& 1& 0& 0 \\
6& 4& 0& 1& 1 
\end{array}&\begin{array}{c}
 -6  \\
 -12 
\end{array}&\begin{array}{cccc}
1& 1& 2& 3\\
1& 5& 3& 7
\end{array}&\begin{array}{cc}
-3 & -4 \\
-9 & -7 
\end{array}\\
\hline
\end{array}\end{aligned}\end{equation} 
 and embed it into a larger GLSM by adding a new gauge field, and fermionic and chiral fields:
\begin{equation}\begin{aligned}&\begin{array}{|c|c| |c|c|}
\hline
x_i & \Gamma^j & \Lambda^a & p_l \\ \noalign{\hrule height 1pt}\begin{array}{cccccccc}
3 & 2 & 1 & 0 & 0 & 0 & 0 & 0 \\
6 & 4 & 0 & 1 & 1 & 0 & 0 & 1 \\
0 & 0 & 0 & 0 & 0 & 1 & 1 & 0
\end{array}&\begin{array}{ccc}
 -6 & 0 & 0 \\
 -12 & 0 & -1 \\
 0 & -1 & 0
\end{array}&\begin{array}{cccccc}
1 & 1 & 2 & 3 & 3 & 3 \\
1 & 5 & 3 & 7 & 8 & 9 \\
0 & 0 & 0 & 0 & 0 & 0
\end{array}&\begin{array}{cccc}
-3 & -4 & -3 & -3 \\
-9 & -7 & -8 & -9 \\
-1 & 0 & 0 & 0
\end{array}\\
\hline
\end{array}\end{aligned}\end{equation} 
After performing the combinatoric ``exchange" (i.e. the usual TSD procedure), this yields the new TSD geometry
 \begin{equation}\begin{aligned}&\begin{array}{|c|c| |c|c|}
\hline
x_i & \Gamma^j & \Lambda^a & p_l \\ \noalign{\hrule height 1pt}\begin{array}{cccccccc}
3 & 2 & 1 & 0 & 0 & 0 & 0 & 0 \\
6 & 4 & 0 & 1 & 1 & 0 & 0 & 1 \\
0 & 0 & 0 & 0 & 0 & 1 & 1 & 0
\end{array}&\begin{array}{ccc}
 -6 & 0 & 0 \\
 -12 & -1 & 0 \\
 0 & -1 & -1
\end{array}&\begin{array}{ccccc}
1 & 1 & 2 & 3 & 3  \\
1 & 5 & 3 & 7 & 8  \\
0 & 0 & 0 & 0 & 1
\end{array}&\begin{array}{ccc}
-3 & -4 & -3\\
-9 & -7 & -8 \\
-1 & 0 & 0
\end{array}\\
\hline
\end{array}\end{aligned}\end{equation} 
 
The advantage of this new example is that, it is possible to compute the spectral cover of both sides easily\footnote{In the previous example the base, $\mathbb{P}^1$, was defined as a conic inside $\mathbb{P}^2$. However, the spectral cover equations would be in terms of the ambient space coordinates and imposing the non linear relations between the coordinates to define the $\mathbb{P}^1$ makes the situation somewhat obscure.} and they are both reducible but still reduced
\begin{eqnarray}
&& S_1 = F_6 (f_2 X + f_6 Z^2), \label{specy1}\\
&& S_2 = F_7 (f_1 X +f_5 Z^2). \label{specy2}
\end{eqnarray} 
 
As in the previous example, we can readily construct the F-theory geometry of both sides, and check whether they are related by exchanging the fibration or not. The Weierstrass polynomials, $F$ and $G$, of the dual F-theories is given by,

\begin{eqnarray}
F_1 = {\cal O}(u_1^5) + u_1^4 u_2^4 f_1^8(v_1,v_2)+u_1^3 u_2^5 F_6(v_1,v_2) f_2(v_1,v_2), \\
G_1 = {\cal O}(u_1^7)  + u_1^6 u_2^6 g_1^{12} (v_1,v_2) +u_1^5 u_2^7 F_6 (v_1,v_2) f_6(v_1,v_2), \\
\\
F_2= {\cal O}(v_1^5)  + v_1^4 v_2^4 f_2^8(u_1,u_2)+v_1^3 v_2^5 F_7(u_1,u_2) f_1(u_1,u_2), \\
G_2 = {\cal O}(v_1^7)   + v_1^6 v_2^6 g_2^{12} (u_1,u_2) +v_1^5 v_2^7 F_7 (u_1,u_2) f_5(u_1,u_2).
\end{eqnarray}

As in the previous example, the change in fibration can be realized in $F_1$ and $G_1$ by re-expanding these polynomials in terms of $v_1$ and $v_2$. Then if the dual (F-theory) geometries are related through changing the fibration, after this rearrangement, $F_1$ and $G_1$ must be equal to $F_2$ and $G_2$. \\

But since $F_2$ and $G_2$ have an order three zero at $v_1=0$, it means that $f_1^8(v_1,v_2)$ and $g_1^{12}(v_1,v_2)$ must have an order three zero at $v_1=0$. Recall that these two polynomials are the $f$ and $g$ of the dual heterotic $K3$ surface, so the above argument tells us if the TSD geometries are related to the different fibrations of the same geometry in F-theory, both TSD Calabi-Yau 2-folds must have an $E_7$ singularity at some point on the base. Thus once again, we see that exchange of fibration leads to a perturbative/non-perturbative duality in heterotic \cite{Morrison:1996pp} and not the apparent correspondence arising from TSD.

In summary, if we start with two perfectly smooth TSD geometries, they cannot be related through different $K3$-fibrations of a single F-theory 3-fold. But if we allow both $K3$ surfaces to be singular, and at the same time put bundles/small instantons over them, they might be dual to a single geometry in F-theory\footnote{But we should recall that the GLSM is only a perturbative formulation and clearly lacks information about the full string theory in such a context.}.

Having determined that the multiple fibrations are not describing the TSD exchange in $6$-dimensions, we can take a step back and ask \emph{what F-theory correspondence is induced by TSD in $6$-dimensions?} Since the spectral covers in \eref{specy1} and \eref{specy2} are relatively simple, we can try to roughly figure out some generalities about the F-theory duals of each of them. Let us start with the first one. The topology of the vector bundle fixes the dimension of the moduli space of the bundle,
\begin{eqnarray}
h^1(V_1 \otimes V_1^*) = 42.
\end{eqnarray}
We can describe them in terms of the spectral data as follows, 
\begin{eqnarray}\label{dimofMV}
dim(\mathcal{M}_V) = dim(cplx(C))  + dim(Jac(C)) + 6 pts + 6\times dim(Jac(E)) + gluing,
\end{eqnarray}
where $C$ is the irreducible smooth curve defined by $f_2 X+f_6 Z^2$, by $6 pts$ we mean the degrees of freedom in choosing the location of the six points defined by the zero set of $F_6=0$, and over them we have 6 elliptic curves (whose Jacobians must also be taken into account), and finally ``gluing" denotes the degrees of freedom associated with the choice of spectral sheaf at the intersection of the 6 vertical fiber with $C$.  
The genus of $C$ can be computed easily,
\begin{eqnarray}
g(C) = 9.
\end{eqnarray} 
Therefore, the dimension of the Jacobian and the complex structure of $C$ must be 9. On the other hand, obviously, Jacobian of $E$ is 1-dimensional, and the contribution of the ``gluing" is 12-dimensional (each vertical fiber intersects $C$ at 2 points). Therefore the total dimension of the Moduli space is,
\begin{eqnarray}
dim(\mathcal{M}_V) = 9+9+6+6+12 =42.
\end{eqnarray} 

Now, to obtain the F-theory EFT we must use the spectral data as explained before, and infer the form of the complex structure of the CY 3-fold. From this procedure it can be seen that there are 6 (4,6,12) points in the F-theory geometry. Since the heterotic dual is a perturbative model, we should consider these singularities as the singular limit of the following deformations,
\begin{eqnarray}
F_1 ={\cal O}(u_1^5) + u_1^4 u_2^4 f_1^8(v_1,v_2)+u_1^3 u_2^5 (F_6(v_1,v_2) f_2(v_1,v_2) +\epsilon F_8(v_1,v_2) ), \\
G_1 = {\cal O}(u_1^7) + u_1^6 u_2^6 g_1^{12} (v_1,v_2) +u_1^5 u_2^7( F_6 (v_1,v_2) f_6(v_1,v_2) + \lambda F_{12}(v_1,v_2)),
\end{eqnarray}
where $\epsilon$ and $\lambda$ correspond to deforming the Higgs field over the 7-branes \cite{Beasley:2008dc, Anderson:2013rka}. Therefore we can deform these two theories into each other by continuously deforming the Higgs bundle. This reflects the fact the moduli space of the vector bundles on $K3$ is connected.  Phrased differently, the existence of apparent $(4,6,12)$ points in the putative dual F-theory indicates that such solutions can \emph{only be dual to the expected perturbative heterotic theories} in the case that T-brane solutions arise. This has been seen before in \cite{Aspinwall:1998he} and is a substantial hint that G-flux must play an important role in the non-trivial F-theory correspondence expected in $4$-dimensional compactifications.

It is worth commenting briefly also on another branch of the theory visible from this singular limit. We can increase the number of tensor multiplets in the $6$-dimensional YM theory by performing small instanton transitions (i.e. moving NS5/M5 branes off the $E_8$ fixed plane in the language of heterotic M-theory). For bundles described as spectral covers, this small instanton limit is visible by the spectral cover becoming reducible and vertical components (corresponding to small instantons) appearing (note that this limit must also set all gluing data to zero). Naively it seems that this limit appears different for the TSD pair of bundles defined by \eref{specy1} and \eref{specy2} since they exhibit different degree polynomials defining their vertical components (i.e. $F_6$ vs. $F_7$). However, this is simply a statement that the mapping of moduli in this case may exchange what are spectral cover deformations in one description with data associated to the Jacobian of the spectral cover (i.e. gluing data in this singular case). To really obtain the same point in moduli space, we must consider a scenario in which both halves of the TSD gain the same number of tensor multiplets (i.e. we pull either 6 or 7 5-branes into the bulk). In this case it would be intriguing to analyze the dual F-theory geometry -- which would correspond to blowing up the base of the elliptic fibration. We expect in this case that the F-theory 3-fold will still be $K3$ fibered but no longer of such a simple form. In particular, the elliptic fibration over a Hirzebruch surface would be modified to become a more general conic bundle \cite{Anderson:2016cdu}. We will return to questions of a similar geometric nature in the following section.

Let us briefly summarize the results of our $6$-dimensional investigation. We have seen that after exchanging $K3$-fibrations within the F-theory geometry, the dual heterotic $K3$ surface must become singular, and therefore perturbative smooth heterotic geometries arising in TSD pairs cannot in general be realized as different fibrations within F-theory. On the other hand we saw that the dual F-theory EFTS arising from the chosen TSD pairs must crucially rely on data from the intermediate Jacobian of the CY 3-fold -- so-called T-brane solutions -- in order to give rise to the same physical theories. Starting from such points we can deform back to smooth points in the CY 3-fold moduli space and identify the theories. Any possible correspondences within the tensor branch of the $6$-dimensional theories must involve more complicated $K3$-fibrations (i.e. conic bundles) and we leave this exploration to future work.

%\subfile{4dftheory}

\section{F-theory duals of $4$-dimensional heterotic TSD pairs}\label{4D_thoughts}
In Section \ref{sec:CICY} we provided a non-trivial example of a heterotic TSD pair in which both $X$ and $\tilde{X}$ were elliptically fibered. It is now natural to ask -- \emph{what are the F-theory duals of these heterotic theories?} As we will explain below, this example (and others like it that we have found) seem to force beyond the arena of ``standard" heterotic/F-theory duality (as in the canonical reference \cite{Friedman:1997yq}) by including several important features in the dual geometries. In this section, we will not try to solve all the obstacles that arise at once. Instead, we will outline what can be determined about the dual F-theory geometries and where new tools will be needed to fully probe this correspondence. Many of these we are currently developing \cite{us_mohsen,us_conicbundles} and we hope to definitively answer these questions in future work.

As a first step towards determining the dual F-theory geometry, the data of the heterotic bundle must be taken through a Fourier-Mukai transform to be presented as spectral data  (see the discussion in Section \ref{6D_Section}). However in this we immediately encounter several problems. The first of these is that unlike in the case of heterotic/F-theory dual pairs studied in the literature to date, neither of these heterotic CY elliptic 3-folds is in Weierstrass form.

To be specific we focus on the examples in Section \ref{sec:CICY} (though similar obstacles will arise in general in this context). Recall that each of the CY 3-folds listed in \eref{4d_dual_cy} admitted two rational sections. Those for $X$ in \eref{4d_dual_cy} lie in the following classes 
 
 \begin{eqnarray}
[ \sigma_1(X) ]= -D_1 + D_2 + D_3, &&\quad [\sigma_2(X)] = 2 D_1 - D_2 + 5 D_3.\nonumber
\end{eqnarray}  
where $D_i$ are a basis of divisors on $X$ (inherited from the ambient space hyperplanes by restriction). By ``rational" it is meant that these divisors are isomorphic to blow ups of the base manifold (in this case $P^2$). The first difficulty with this example is that the standard Fourier-Mukai transformation with Poincare bundle is not applicable here. The reason is we need the zero section to intersect exactly one point on \textit{every fiber}, but both of the sections described above wrap around a finite number of rational curves (which are components of reducible fibers). We have shown \cite{us_mohsen} that in specific situations one can use flop transitions to make one of the sections holomorphic, and since derived categories are invariant under the flop transitions (i.e. there is a specific Fourier-Mukai functor for flops), it is still possible to define the spectral data in the ``flopped" geometry. However the example given in \eref{4d_dual_cy} proves to be too complicated to be analyzed in this manner since $\sigma_1$ and $\sigma_2$ wrap around 27 and 127 rational curves respectively, rendering the necessary birational transformations (i.e. flops) impractical. 

In principle, one might hope to bypass this difficulty by transitioning $X$ directly to its Weierstrass form (by blowing down the reducible components of fibers), following the Deligne procedure outlined in \cite{Anderson:2016ler,Anderson:2016cdu}. However, this poses difficulties in a heterotic theory in that it is unclear how the heterotic bundle data should be appropriately mapped to this singular limit of $X$.

None-the-less, if we choose $\sigma_1$ as the zero section, it can be demonstrated that the spectral cover in the singular Weierstrass limit, has the same divisor class as before (this is seen by taking the FM transform before blowing down the reducible fiber components). In other words, if we write the second Chern class as
\begin{eqnarray}
c_2(V) = 36 \sigma_1 H + 14 S_{sh} H +156 f, 
\end{eqnarray}
where $H$ is the (pull-back of the) hyperplane divisor in the base, $S_{sh}$ is the divisor corresponding to the Shioda map \cite{shioda,shioda2,COM:213767} for non trivial Mordell-Weil group,
\beq
S_{sh}= \sigma_2-\sigma_1 - 18 H
\eeq
 and $f$ is the fiber class. In terms of these divisors, the class of the spectral cover, $S$, in the singular limit will be,
\begin{eqnarray}\label{spec_class}
[S] = 6 \sigma_1 +36 H.
\end{eqnarray}

We might hope to get some information about the F-theory geometry just from the spectral cover alone. Naively, we may write the algebraic formula for the spectral cover whose class is given in \eref{spec_class} as
\begin{eqnarray}\label{spec_eqn}
S= f_{36} z^3+f_{30} x z+ f_{27} y. 
\end{eqnarray}
where $f_i$ are generic polynomials of degree $i$ over $\mathbb{P}^2$. A generic deformation of the spectral cover of the form \eref{spec_eqn} can be obtained by counting the degrees of freedom in the  polynomials $f_{36}, f_{30}, f_{27}$ which contain 703, 496 and 406 parameters, respectively. Immediately we see that these numbers much higher than the dimension of the vector bundle moduli space in our example, which is 292-dimensional. Thus, we can see that the FM-transform of the monad in \eref{eq:eg31} is certainly \emph{not} a generic spectral cover. This is not too surprising. We have seen examples of the spectral cover of monads in Section \ref{6D_Section} and there it was clear that the polynomials are not generic, rather they are dictated by the monad's map (see also \cite{Ber,Donagi:2011dv}). In principle, a similar story happens in the current case. We expect that the spectral cover may also be non-reduced/reducible \cite{Ber}. However, regardless of its explicit form, the question arises, why is the spectral cover forbidden from assuming a generic form? That is, given an explicit starting point (in which the polynomials are determined by the monad map as in \eref{eq:eg31}) why is the deformation space restricted? 

We expect that the answer to this lies with the other half of the spectral data of this monad, that is, the rank 1 sheaf \cite{Friedman:1997yq} supported over the spectral cover in \eref{spec_class} and \eref{spec_eqn}. It has been observed previously \cite{Donagi:2009ra} that the Picard group of $S$ may ``jump" at higher co-dimensional loci in moduli space -- i.e. so-called Noether-Lefschetz loci. This phenomenon could ``freeze" the moduli of the spectral cover to a sub-space compatible with the form of the monad map (see also \cite{Anderson:2016kuf}). In terms of the $4$-dimensional, ${\cal N}=1$ EFT, the reduction in the apparent number of singlets (i.e. the non generic form of the spectral cover) is a symptom of existence of a specific superpotential -- arising from the Gukov-Vafa-Witen form \cite{Gukov:1999ya}:
\beq
W \sim \int_X H \wedge \Omega
\eeq
where $H \sim dB + \omega_3^{YM} - \omega_3^{Lorentz}$, and $\omega_3 =F \wedge A - \frac{1}{3} A \wedge A \wedge A$ is the Chern Simons 3-form (and the associated Lorentz quantity built from the spin connection in $\omega_3^{Lorentz}$ and $\Omega$ is the holomorphic $(0,3)$ form on $X$. The existence of this superpotential arises from the presence of the gauge bundle (rather than from quantized flux) (see \cite{Donagi:2009ra,Anderson:2011ty,Anderson:2010mh} for related discussions) but none-the-less stabilizes vector bundle moduli.

As a result, in the dual F-theory EFT, we also expect the existence of a superpotential. Geometrically, since the spectral cover determines part of the complex structure moduli of the Calabi-Yau fourfold, it is clear that the dual of the bundle data given in \eref{eq:eg31} should include a specific G-flux that stabilizes the moduli through the GVW superpotential. It should be noted that there is another way to see the requirement for this flux: since there are no $D3$ branes in the F-theory dual (since we have chosen $c_2(X)=c_2(V)$ in the heterotic theory), G-flux is also necessary for anomaly cancellation. 

Although we have not yet explicitly calculated the FM transform of the heterotic bundle or determined the dual F-theory geometry. The arguments above show that whatever the F-theory geometry, G-flux must play a prominent role and therefore it cannot be ignored. A similar set of arguments can also be made about the F-theory dual of the heterotic TSD geometry $(\tilde{X},\tilde{V})$. In this case as well, the naive deformations of the spectral cover are much larger than the dimension of the vector bundle moduli space, and therefore we conclude that Noether-Lefschetz loci/G-flux should be in play.

Despite the fact that flux must be involved in the putative F-theory duality, it still remains to be asked whether the dual F-theory 4-folds might still exhibit multiple fibration structure? That is could the geometric scenario described in the introduction with these compatible elliptic/$\mathbb{P}^1$ (and hence $K3$) fibrations exist?
\begin{displaymath}
 \xymatrix{
& Y_4 \ar[d]^{\mathbb E}_{\rho_f} & \\
& B_3 \ar[ld]_{\mathbb P^1}^{\sigma_f} \ar[rd]^{\mathbb P^1}_{\tilde \sigma_f} &\\
B_2& & \widetilde B_2}
\end{displaymath}
On this front, once again we see that we must quickly leave behind the ``standard" geometry of heterotic/F-theory duality. As reviewed in Section \ref{Fth}, if the heterotic CY 3-fold is in Weierstrass form, the construction of \cite{Friedman:1997yq} generates a threefold base, $B_3$ (see \eref{p1bund}) for the CY 4-fold that is a $\mathbb{P}^1$-bundle over the base $B_2$ (which is the base of the dual elliptically fibered CY 3-fold and $K3$-fibered 4-fold). The topology of this bundle (i.e. $B_3$ itself) is determined by the second Chern class of the heterotic bundle $c_2(V)$. In this context then, we can ask whether or not such a base could admit two different descriptions as a $\mathbb{P}^1$ bundle? While multiply fibered $\mathbb{P}^1$ bundles certainly exist (for example the ``generalized Hirzebruch" toric 3-fold defined as the zero twist over $\mathbb{F}_n$ or the $n$-twist over $\mathbb{F}_0$ \cite{Berglund:1998ej,Anderson:2016cdu}), it is easy to demonstrate that 
\beq
h^{1,1}(B_3)=1+h^{1,1}(B_2)
\eeq
for any $\mathbb{P}^1$ bundle. As a result, it is clear that \emph{there exists no multiply fibered $\mathbb{P}^1$-bundles compatible with the $B_2$ and $\tilde{B}_2$ arising in Section \ref{sec:CICY}} since for those manifolds $B_2=\mathbb{P}^2$ and $\tilde{B}_2=dP_1$. Hence $h^{1,1}(B_2)<h^{1,1}(\tilde{B}_2$ and $h^{1,1}(B_3) \neq h^{1,1}(\tilde{B_3})$ for 3-fold bases constructed as $\mathbb{P}^1$-bundles. 

From the results above we would be tempted to conclude that the hypothesis we set out to test in the literature (i.e. is heterotic TSD dual to multiply fibered geometries in F-theory?) is manifestly false. However, we must first recall that the construction of $K3$ fibrations in terms of $\mathbb{P}^1$-bundle bases $B_3$ as commonly used in the literature is not the only possible structure. More general 3-fold bases, $B_3$ are possible which are $\mathbb{P}^1$ fibrations \emph{but not $\mathbb{P}^1$ bundles}. These fibrations degenerate (as multiple $\mathbb{P}^1$s) over higher co-dimensional loci in the base $B_2$ and are known as ``conic bundles" in the literature (see e.g. \cite{sarkisov}).

If we consider this more general class of bases for CY 4-folds it seems that some possibilities remain. For example, the following threefold
\begin{equation}\label{conic}
B_3 = \left[ \begin{array}{c|cc} 
\IP^2 & 0&1\\  \IP^1 & 1&0\\ \IP^2 &1&1\end{array} \right]
\end{equation}
is manifestly fibered over both $\mathbb{P}^2$ and $dP_1$ as required. However, it is unclear that the generic ``twist" of such a fibration is compatible with the topology of the bundles defined in Section \ref{sec:CICY}.
It is possible to generalize simple constructions like the one above to accommodate more general twists by choosing more general toric ambient spaces. However, in each case we hit a new problem in that the stable degeneration limits of $\mathbb{P}^1$ bundles such as that in \eref{conic} are not yet understood in the literature (though we are considering such geometries in separate work \cite{us_conicbundles}). As a result, it is a non-trivial task to determine whether such a geometry might arise in the F-theory duals of the examples outlined in Section \ref{sec:CICY}. To check this we need precise spectral data. But as explained before, finding the Fourier-Mukai transforms of the heterotic bundles, while possible in principle, is beyond our current computational limits for the bundles in Section \ref{sec:CICY}. 

For now though, we can conclude that whatever the F-theory correspondence induced from $(0,2)$ target space duality may be, it must expand the current understanding of heterotic/F-theory duality both via the crucial inclusion of G-fluxes (including possibly limits and T-brane solutions) and via more general geometry -- in particular $K3$/$\mathbb{P}^1$-fibrations -- than has previously been considered.

%\subfile{conclusions}

\section{Conclusions and future directions}

In this work we have taken a first step towards exploring the consequences of $(0,2)$ target space duality for heterotic/F-theory duality. In an important proof of principle, we have illustrated that heterotic TSD pairs exist in which both halves of the geometry exhibit Calabi-Yau threefolds with elliptic fibrations. As a result, it is clear that some F-theory correspondence should be induced in these cases. We take several steps to explore the properties of this putative duality. First, we consider the conjecture made previously in the literature that the F-theory realization of TSD could be multiple $K3$ fibrations of the same elliptically fibered Calabi-Yau $4$-fold background of F-theory. To explore this possibility in earnest, we begin in $6$-dimensional compactifications of heterotic string theory/F-theory and demonstrate that in general multiple fibrations within F-theory CY backgrounds cannot correspond to the (topologically trivial) TSD realizable for bundles on $K3$ surfaces. Finally, we provide a sketch of the open questions that arise when attempting to directly compute the F-theory duals of $4$-dimensional heterotic TSD geometries. In particular, we demonstrate that multiple $K3$ fibrations in F-theory cannot account for $(0,2)$ TSD in the case that the threefold base, $B_3$ of the F-theory elliptic fibration takes the form normally assumed -- that of a $\mathbb{P}^1$ bundle over a two (complex) dimensional surface, $B_2$.

There are a number of future directions that naturally lead on from this study, most importantly to explicitly determine the F-theory mechanism that generates dual theories from potentially disparate $4$-fold geometries. We hope to understand this correspondence in future work. The present study has shed light on these questions however, and highlighted areas where the current state-of-the art in the literature is insufficient to determine the dual heterotic/F-theory geometries. 

As noted in Section \ref{4D_thoughts}, it is clear that
new tools will be needed to fully determine this duality. The new geometric features that must be understood in heterotic/F-theory duality in this context clearly extend beyond the canonical assumptions made in \cite{Friedman:1997yq} and new tools must be developed. These include the following open problems in heterotic/F-theory duality
\begin{itemize}
\item Heterotic compactifications on elliptic threefolds with higher rank Mordell-Weil group (as in the examples in Section \ref{sec:CICY}).
\item F-theory compactifications on threefold bases that are $\mathbb{P}^1$ fibered, but not $\mathbb{P}^1$ bundles. I.e. F-theory on elliptic fibrations with \emph{conic bundle} (see e.g. \cite{sarkisov}) bases.
\item F-theory duals of degenerate (i.e. non-reduced and reducible) heterotic spectral covers. These seem to be a ubiquitous feature in the  context of $(0,2)$ target space duality since the spectral data of monad bundles appear to be generically singular \cite{Ber}.
\item $4$-dimensional T-brane solutions of F-theory (expected to arise in the context of degenerate spectral covers above \cite{Anderson:2013rka,Anderson:2017rpr,Anderson:2017zfm}).
\end{itemize}
This last point seems to be an essential part of the story for $4$-dimensional heterotic/F-theory pairs since degenerate spectral data naturally arise for monad bundles (and hence geometries arising from $(0,2)$ GLSMs). Moreover, the arguments of Section \ref{4D_thoughts} make it clear that the degrees of freedom of an expected dual F-theory $4$-fold must be constrained by flux in order to match the moduli count of the heterotic theory. Several of these ``missing ingredients" are currently being studied (see \cite{us_mohsen} for generalizations of heterotic geometries in heterotic/F-theory duality and \cite{us_conicbundles} for a study of F-theory on conic bundles). We hope that the present work illustrates the need for these new tools and demonstrates that there remain many interesting open questions within the context of $4$-dimensional heterotic/F-theory duality. We will return to these open questions in future work.

\section*{Acknowledgements}
The authors would like to thank James Gray, Paul Oehlmann and Nikhil Raghuram for useful discussions. In addition, LA and MK gratefully acknowledge the hospitality of the Simons Center for Geometry and Physics (and the semester long program, \emph{The Geometry and Physics of Hitchin Systems}) during the completion of this work. The work of LA is supported by NSF grant PHY-1720321.

%\subfile{eg_appendix}

\appendix

\section{Appendix A}

In this appendix we present some of the exotic cases we encountered during the search for finding ``good examples"  of stable, smooth vector bundles over bases that are Weierstrass elliptic fibrations. All of these examples pass the usual necessary conditions for stability such as $h^0(V)=0$ and Bogomolov topological constraint, but either the spectrum charged hypermultiplets of the 4d effective are different or the total moduli is not conserved. By using careful Fourier-Mukai analysis we can show that the first example is indeed unstable, so it explains the discrepancy, but the other two are perfectly stable vector bundles, and we are unable to explain the reason. In the third example that spectrum match on both sides one may suggest that the existence of the flux (which must exist due to the generically non-reduced spectral cover) may stabilize the moduli space.

\begin{equation}\begin{aligned}&\begin{array}{|c|c| |c|c|}
\hline
x_i & \Gamma^j & \Lambda^a & p_l \\ \noalign{\hrule height 1pt}\begin{array}{ccccccc}
3& 2& 1& 0& 0& 0& 0  \\
0& 0& -2& 1& 1& 0& 0  \\
0& 0& -2& 0& 0& 1& 1 
\end{array}&\begin{array}{c}
 -6  \\
 0  \\
 0 
\end{array}&\begin{array}{ccccc}
1& 0& 1& 1 & 1 \\
1& 3& -2& 2 & 1 \\
0& 3& -2& 3 & 0 
\end{array}&\begin{array}{cc}
 -1& -3 \\
 -1& -4 \\
 -1& -3 
\end{array}\\
\hline
\end{array}\end{aligned}\end{equation}
 with the second Chern classes as:
\begin{eqnarray}
c_2(X)&=&11 \sigma^2 + 2 \sigma D_1+ 2\sigma D_2 - 3 D_1^2- 4 D_1 D_2 - 3 D_2^2 = 24 \sigma D_1+24 \sigma D_2-4 D_1 D_2, \nonumber\\
c_2(V)&=&3 \sigma^2 + 11\sigma D_1+ 9\sigma D_2 -  D_1^2- 6 D_1 D_2 - 6 D_2^2 = 17 \sigma D_1+15 \sigma D_2-6 D_1 D_2 \nonumber,
\end{eqnarray} 
where $\sigma$, $D_1$, and $D_2$ are the section and base divisors correspondingly, with $D_1^2=D_2^2=0$, and $D_1 D_2 = f$ the class of the generic fiber $f$. The anomaly cancellation is not satisfied in the strong sense, but  we can still make sense of it at least as heterotic string theory  (may be not GLSM, but well defined as heterotic string theory).
Again we embed this GLSM into a larger one:
\begin{equation}\begin{aligned}&\begin{array}{|c|c| |c|c|}
\hline
x_i & \Gamma^j & \Lambda^a & p_l \\ \noalign{\hrule height 1pt}\begin{array}{ccccccccc}
3& 2& 1& 0& 0& 0& 0& 1& 0  \\
0& 0& -2& 1& 1& 0& 0& -3& 1  \\
0& 0& -2& 0& 0& 1& 1& -2& 1 
\end{array}&\begin{array}{ccc}
 -6& -1& 0  \\
 0& 3& -1  \\
 0& 2& -1 
\end{array}&\begin{array}{ccccc}
1& 0& 1& 1& 1   \\
1& 3& -2& 2& 1  \\
0& 3& -2& 3& 0 
\end{array}&\begin{array}{cc}
 -1& -3  \\
 -1& -4 \\
 -1& -3 
\end{array}\\
\hline
\end{array}\end{aligned}\end{equation}
 with the degrees of the monad maps is as follows:
\begin{equation}
\begin{array}{|c|c|}
\hline
F^1 & F^2 \\
\hline
\begin{array}{ccccc}
0 & 1 & 0 & 0 & 0 \\
0 & -2 & 3 & -1 & 0 \\
1 & -2 & 3 & -2 & 1 
\end{array} & \begin{array}{ccccc}
2 & 3 & 2 & 2 & 2 \\  
3 & 1 & 6 & 2 & 3 \\ 
3 & 0 & 5 & 0 & 3 
\end{array} \\
\hline
\end{array}
\end{equation}
After exchanging $\Gamma^{2}$, $\Gamma^{3}$ (degree $||G_{2}||$,  $||G_{3}||$respectively) with $F^1_{1}$, $F^1_{2}$  respectively, and integrating out the repeated entries, the dual
$(\widetilde X, \widetilde V)$ can be written as follows:
\begin{equation}\begin{aligned}&\begin{array}{|c|c| |c|c|}
\hline
x_i & \Gamma^j & \Lambda^a & p_l \\ \noalign{\hrule height 1pt}\begin{array}{ccccccc}
3& 2& 1& 0& 0& 0& 0  \\
0& 0& -3& 1& 1& 1& 0  \\
0& 0& -2& 0& 0& 1& 1 
\end{array}&\begin{array}{c}
 -6  \\
 0  \\
 0 
\end{array}&\begin{array}{ccccc}
1& 0& 1& 1 & 1 \\
0& 4& -2& 2 & 1 \\
0& 3& -2& 3 & 0 
\end{array}&\begin{array}{cc}
 -1& -3 \\
 -1& -4 \\
 -1& -3 
\end{array}\\
\hline
\end{array}\end{aligned}\end{equation}
    
The dual geometry is perfectly smooth and anomaly cancellation condition can also be make sense as before. However, the spectrum of charged scalers are not the same  (i.e, $h^1(V) = 121$ while $h^1(\widetilde V) = 101$.  So there should be a problem. We can argue that it is related to the stability. 

 After a detailed calculation of Fourier-Mukai transform of V, it becomes clear that $FM^1(V)$ is of relative rank 1 and degree 2. On the other hand $FM^0(V)$ is also non zero with relative rank and degree 1 and -1. It is well known that Fourier-Mukai transformation of a sheaf of relative rank and degree $(n,d)$ is \textit{complex} of relative rank and degree $(d,-n)$. So it is clear from the above data that the restriction of $V$ on a generic elliptic fiber $E$ is roughly of the form $\mathcal{O}_E (\sigma) \oplus \mathcal{V}_2$, where $\mathcal{V}_2$ is a rank 2 irreducible bundle of degree -1 on $E$. Obviously it tells us that the bundle must be unstable because it is unstable on generic fibers (even though it seems $h^0(V)=0$). As a sanity check we can compute the rank of $\pi_* V$ and $\pi_* (V\otimes \mathcal{O}_X(\sigma))$, and they are 1 and 3 respectively. This is consistent because $h^* (\mathcal{V}_2) = (0,1)$ \footnote{One can also get the same numbers from semi stable bundles with rank 3 and degree zero, so they are just necessary conditions.}. A similar statement can be made about the TSD set up.

\section{Appendix B} 
 
In this appendix, we present an example of a TSD pair in which the geometries $(X,V)$ and $(\tilde{X},\tilde{V})$ are actually equivalent geometries, even though they are described by different algebraic descriptions (of manifolds and monad bundles). Another interesting feature in this case is that both sides of this ``trivial" correspondence are elliptically fibered, however the base manifolds are two \emph{different} Hirzebruch surfaces, $\mathbb{F}_0$ and $\mathbb{F}_2$. These base surfaces are distinct as complex manifolds but identical as real (and the elliptic CY 3-fold over these different surfaces is the \emph{same} complex manifold).  This demonstrates that even ``trivial" TSD correspondences may involve interesting geometric structure. \\

In the following example the bundle $\widetilde V$ on $\widetilde X$ as a non-trivial rewriting of  bundle $V$ on $X$. Both of the CY 3-folds are  weighted projective space $\IP^2 [123]$ fibered Calabi-Yau 3-folds.  For $X$ the base is Hirzebruch surface $\mathbb{F}_0$, i.e, $B_2$ = $\mathbb P^3 [2]$ while for $\widetilde X$ the base is  $\widetilde {B_2} = \left[ \begin{array}{c|cc} \IP^3 & 1&1\\ \IP^1 & 1&1\end{array} \right]$, which is generically $\mathbb{F}_0$ but at special complex structure moduli it ``jumps"
 to become $\mathbb{F}_2$ \cite{Hubsch:1992nu}. A $(0,2)$ target space map can be found that takes $X$ to $\widetilde X$ (this can be achieved by adding a $\mathbb P^1$ to the configuration as usual for a $U(1)$-changing TSD pair). On this manifold, both tangent bundle and non-tangent bundle will be studied.

\subsection{Non trivial rewriting with tangent bundle}

Let us first consider the case of a deformation of the tangent bundle. The GLSM charge matrix is a general deformation of $(X,V=TX+{\cal O}^{\oplus 2})$ and can be written as follows:
\begin{equation}\begin{aligned}&\begin{array}{|c|c| |c|c|}
\hline
x_i & \Gamma^j & \Lambda^a & p_l \\ \noalign{\hrule height 1pt}\begin{array}{ccccccc}
3& 2& 1& 0& 0& 0& 0  \\
0& 0& -2& 1& 1& 1& 1  
\end{array}&\begin{array}{cc}
 -6 & 0 \\
 0  & -2
\end{array}&\begin{array}{ccccccc}
3& 2& 1& 0& 0& 0& 0  \\
0& 0& -2& 1& 1& 1& 1  
\end{array}&\begin{array}{cc}
 -6 & 0 \\
 0  & -2
\end{array}\\
\hline
\end{array}\end{aligned}\end{equation}
Following the procedure described in previous section,
we will end up with the new charge matrix of the target space dual $(\widetilde X, \widetilde V)$ as:
\begin{equation}\begin{aligned}&\begin{array}{|c|c| |c|c|}
\hline
x_i & \Gamma^j & \Lambda^a & p_l \\ \noalign{\hrule height 1pt}\begin{array}{ccccccccc}
3& 2& 1& 0& 0& 0& 0 & 0& 0  \\
0& 0& -2& 1& 1& 1& 1& 0& 0 \\
0& 0&  0& 0& 0& 0& 0& 1& 1 
\end{array}&\begin{array}{ccc}
 -6 & 0& 0 \\
 0  & -1& -1 \\
  0  & -1& -1
\end{array}&\begin{array}{ccccccc}
3& 2& 1& 0& 0 & 0& 0  \\
0& 0& -2& 1& 1& 0& 2 \\
0& 0&  0& 0& 0& 1& 0 
\end{array}&\begin{array}{cc}
 -6 & 0 \\
 0  & -2 \\
  0  & -1
\end{array}\\
\hline
\end{array}\end{aligned}\end{equation}

The number of both charged and uncharged geometric moduli of the theories on these two manifolds are the same, which suggests that 
 they are indeed target space dual to each other. Such degree of freedom counting is given by:
 %(NEEDS DOUBLE CHECK of $h^1(End_0(V))$)
\begin{eqnarray}\nonumber
h^{*}(V)=(0,241,1,0)\quad h^{1,1}(X)+h^{2,1}(X)+h^1(End_0(V))=3+243+1074=1320 \\
h^{*}(\widetilde{V})=(0,241,1,0)\quad h^{1,1}(\widetilde{X})+h^{2,1}(\widetilde{X})+h^1(End_0(\widetilde{V}))=3+243+1074=1320
\end{eqnarray}

\subsubsection*{Calculate twist of $V$ and $\widetilde V$}

Starting with heterotic theory, without loss of generality,  the second chern class can be  splits as (\ref{eq:c_2})
 and the heterotic Bianchi identity will implies further that $\eta$ can be parameterized as (\ref{eq:twist})
with the twist of the theory $T' = T$.
 %$V$ is deformation of tangent bundle, $\widetilde V$ is also similar to deformation of tangent bundle. 
 In order to get the twist in our example, one can first calculate the second chern class of  $V$ and $\widetilde V$ as
\bea
&c_2(V) = c_2(TX) = 11J_1^2+2J_1J_2 -2J_2^2 \\
&c_2(\widetilde V) = c_2(\widetilde{TX}) = 11J_1^2+2J_1J_2 -3J_2^2+2J_2J_3.
\label{c22}
\eea
For both $X$ and $\widetilde X$, the section can be parameterized as $\sigma = J_1- 2 J_2$, and the section satisfy the birational condition $\sigma^2 = - c_1 (B) \sigma$. Then by applying eq.(\ref{eq:c_2}, \ref{eq:twist}) we get:
\bea
\eta = 24 J_2, \quad T= 12 J_2 = 6 c_1 (B)
\eea
for both $V$ and $\widetilde V$. This indicates that if we start from a deformation of the tangent bundle, after target space dual we will at least end up with a TSD bundle over the same manifold that is topologically equivalent. %This doesn't only hold in DOF count, but also hold in terms of twist.

\subsubsection*{Complex deformation of  Bundle Moduli}
We can further compare $V$ and  $\tilde V$  by analyzing the deformation of these vector bundles.
Consider the difference of $V$ and $\widetilde V$ defined on $B_2$ and $\widetilde B_2$  in the sequence separately, they are:
\begin{eqnarray}\nonumber
&0 \to V \to \mathcal O(0,1)^{\oplus 2} \to \mathcal O(0,2) \to 0& \\
&0 \to \widetilde V \to \mathcal O(0,0,1) \oplus \mathcal O(0,2,0) \to \mathcal O(0,2,1) \to 0&,
\end{eqnarray}
where $V$ is the kernel of map with two degree $||1||$ polynomial on $B_2 = \left[ \begin{array}{c|c}\IP^3 & 2\end{array}\right]$, $\widetilde V$ is kernel of map $F$ with degree $|| 0,1 || $ and $ ||2,0||$ on $\widetilde {B_2} =  \left[ \begin{array}{c|cc}  \IP^3 &1&1\\ \IP^1 & 1&1\end{array} \right]$.  However  for $(\widetilde B_2, \widetilde V)$, if we first solve the polynomial of degree  $|| 0,1 || $  and put the constraint on the second map with degree $ || 2,0 ||$, the second map will exactly reduce to a degree $|| 2 ||$ polynomial on the manifold $\left[ \begin{array}{c|cc}\IP^3 & 1 & 1\end{array}\right]$.  So it seems that the bundle moduli in $(\widetilde B_2, \widetilde V)$ are transformed to the complex moduli in $(B_2, V)$. Then one would be interesting to ask 
%For $(B_2,V_1)$ and $(\widetilde {B_2}, \widetilde {V_1})$, $\mathbb P^3 [2 | 1 1]$ is the same as $\left[ \begin{array}{cc|cc} 1&1&2&0\\1&1&0&1\end{array} \right]$,
%because when applying the (0,1) constraint to the latter, it becomes $\mathbb P^3 [1 1 | 2]$. {\color{blue}So the constraints from $X$ match $\widetilde V$, and constraints from $\widetilde X$ match $V$.}
%This leads us to the question, 
whether it is possible that the complex structure and bundle moduli exchange in $(X, V)$ and $(\widetilde X, \widetilde V)$.

Before  answering this question, there is an important observation as
$\widetilde {B_2}$ is generically $\mathbb{F}_0$, but at a special point it ``jumps"  to become $\mathbb{F}_2$. 
%We just need to find out what happens at the point where $\widetilde {B_2}$ becomes $F_2$.
Write down the defining equations for $\widetilde {B_2} = \left[ \begin{array}{c|cc}  \IP^3 &1&1\\ \IP^1 & 1&1\end{array} \right]$ as:
\begin{eqnarray}\nonumber
z_0 w_0 + \qquad \qquad\qquad  z_1w_1 = 0 \\
z_2 w_0 +(\sum_{i=0}^{2}a_iz_i+\epsilon z_3)w_1 =0,
\end{eqnarray}
with $[z_0,z_1,z_2,z_3] \in \IP^3$ and $[w_0,w_1] \in \IP^1$. If $\epsilon \neq 0$, this system defines $\mathbb{F}_0$. When $\epsilon = 0$, a $\mathbb P^1$ blows up at $(0,0,0,1) \in \IP^3$, which makes  it becomes  $\mathbb{F}_2$.  So the question about 
whether the complex structure and bundle moduli exchange in $(B_2, V)$ and $(\widetilde B_2, \widetilde V)$ changes to what happens for the geometric moduli of $(\widetilde B_2,  \widetilde V)$ when $\widetilde {B_2}$ becomes $\mathbb{F}_2$, and the same for tuning the map of bundle in the $(B_2, V)$ system.

So in calculating the line bundle cohomology in the new system $(\widetilde B_2, \widetilde V)$,  we will not only count the dimension of the cohomology group appearing in the sequence but also their polynomial representations and the explicit map. More specifically,  we will set $z_3 \in \IP^3$ in our calculation to be zero to deform the $\widetilde B_2$ to be $F_2$ and  see what happens. In this case, the lindbundle cohomology are $h^*(\cO(0,1))=\{2,0,0\}, h^*(\cO(2,0))=\{9,0,0\}, h^*(\cO(1,1))=\{12,0,0\}$ with and without turning the base manifold. Furthermore,  we can check that  the cohomology of bundle $h^*(\widetilde B_2, \widetilde V)=\{4, 5, 0\}$ will not be effected by the  tuning. On the other hand,  we can also tune the complex structure of the map ($x_7=0$ in $F$) in defining the map of $V$ in the  $( B_2,  V)$ system. Again the deformation of map does not change the bundle valued cohomology $h^*(B_2,V) = \{1,2,0\}$.

%{\color{blue}Now we try to tune the map:} For $(\widetilde X, \widetilde V)$, when calculating line bundle cohomology, take the representative polynomials in Koszul sequence, and set the terms depending on z3(i.e. x7) to zero. {\color{blue} It turns out the cohomology doesn't change.} In fact, some line bundle cohomology will change in this process, but none of them is in our case.

%For $(X, V)$, we want to tune the bundle instead of the manifold. Here we set z3(i.e. x7) to zero in the defining map of V. {\color{blue} Again the cohomology doesn't change.}

%{\color{red}CODE OR SCREENSHOT GOES HERE...}

\subsection{Non trivial rewriting  with general vector bundle}

Similarly,  we can consider  another example with the same manifolds but different bundles. 
Again, we start from the following manifold with charge matrix fo the form (X,V):

\begin{equation}\begin{aligned}&\begin{array}{|c|c| |c|c|}
\hline
x_i & \Gamma^j & \Lambda^a & p_l \\ \noalign{\hrule height 1pt}\begin{array}{ccccccc}
3& 2& 1& 0& 0& 0& 0  \\
0& 0& -2& 1& 1& 1& 1  
\end{array}&\begin{array}{cc}
 -6 & 0 \\
 0  & -2
\end{array}&\begin{array}{ccccc}
4& 2& 0& 0& 0  \\
0& -2& 2& 2& 1  
\end{array}&\begin{array}{ccc}
 -6 & 0 \\
 0  & -3
\end{array}\\
\hline
\end{array}\end{aligned}\end{equation}
The second Chern class of  $(X, TX)$ and$(X,V)$ are given by
\begin{eqnarray}\nonumber
&c_2(TX) = 11J_1^2+2J_1J_2 -2J_2^2&\\
&c_2(V) = 8J_1^2+4J_1J_2 -2J_2^2&
\end{eqnarray}
which satisfy the $c_2$ condition $c_2(V) \leq c_2(TX)$.  The target space dual of this theory is given by the form of 
$(\widetilde X, \widetilde V)$:
\begin{equation}\begin{aligned}&\begin{array}{|c|c| |c|c|}
\hline
x_i & \Gamma^j & \Lambda^a & p_l \\ \noalign{\hrule height 1pt}\begin{array}{ccccccccc}
3& 2& 1& 0& 0& 0& 0 & 0& 0  \\
0& 0& -2& 1& 1& 1& 1& 0& 0 \\
0& 0&  0& 0& 0& 0& 0& 1& 1 
\end{array}&\begin{array}{ccc}
 -6 & 0& 0 \\
 0  & -1& -1 \\
  0  & -1& -1
\end{array}&\begin{array}{ccccc}
4& 2& 0& 0& 0  \\
0& -2& 1& 3& 1 \\
0& 0& 1& 0& 0 
\end{array}&\begin{array}{ccc}
 -6 & 0 \\
 0  & -3 \\
 0 & -1
\end{array}\\
\hline
\end{array}\end{aligned}\end{equation}
with second Chern class as:
\begin{eqnarray}\nonumber
&c_2(\widetilde{TX}) = 11J_1^2+2J_1J_2 -3J_2^2+2J_2J_3 &\\
&c_2(\widetilde V) = 8J_1^2+4J_1J_2 -3J_2^2+2J_2J_3&
\end{eqnarray}

Once again we get their twists of the base to be the same:
\begin{eqnarray}\nonumber
\eta = 20J_2 \quad T = 8J_2 \qquad 
\end{eqnarray}
for both $V$ and  $\widetilde V$.
These result indicates that this target space dual is just a kind of rewriting of the origin $(X, V)$.

%%%%%%%%%%%%%%%%%%%%%%%%%%%%%%%%%%%%%%%%%%%%%%%%%%

\end{document}